\documentclass[onecolumn,useAMS,usenatbib]{mn2e}
\usepackage{amsmath}
\usepackage{graphicx}
\usepackage{wasysym}

\title[Core-mantle interactions for Mercury]{Core-mantle interactions for Mercury}
\author[B. Noyelles, J. Dufey and A. Lemaitre]{B. Noyelles$^{1,2}$\thanks{E-mail:
noyelles@imcce.fr}, J. Dufey$^{1}$ and A. Lemaitre$^{1}$ \\
$^{1}$FUNDP (University of Namur) -- Department of Mathematics -- Rempart de la Vierge 8 -- B-5000 Namur -- Belgium \\
$^{2}$ IMCCE (Paris Observatory - USTL - UPMC) -- CNRS UMR 8028 -- 77 avenue Denfert-Rochereau --  75014 Paris -- France }
\begin{document}

\date{Accepted . Received ; in original form }

\pagerange{\pageref{firstpage}--\pageref{lastpage}} \pubyear{xxxx}

\maketitle

\label{firstpage}

\begin{abstract}

Mercury is the target of two space missions: MESSENGER (NASA) which orbit insertion is planned for March 2011, and ESA/JAXA BepiColombo, that should be launched in 2014. Their instruments will observe the surface of the planet with a high accuracy (about 1 arcsec for BepiColombo), what motivates studying its rotation. Mercury is assumed to be composed of a rigid mantle and an at least partially molten core. We here study the influence of the core-mantle interactions on the rotation perturbed by the solar gravitational interaction, by modeling the core as an ellipsoidal cavity filled with inviscid fluid of constant uniform density and vorticity. We use both analytical (Lie transforms) and numerical tools to study this rotation, with different shapes of the core. We express in particular the proper frequencies of the system, because they characterize the response of Mercury to the different solicitations, due to the orbital motion of Mercury around the Sun. We show that, contrary to its size, the shape of the core cannot be determined from observations of either longitudinal or polar motions. However, we highlight the strong influence of a resonance between the proper frequency of the core and the spin of Mercury that raises the velocity field inside the core. We show that the key parameter is the polar flattening of the core. This effect cannot be directly derived from observations of the surface of Mercury, but we cannot exclude the possibility of an indirect detection by measuring the magnetic field.

\end{abstract}


\section{Introduction}

\par Mercury is the target of two current space missions (see e.g. \cite{msgnm04}). The first one, MESSENGER (NASA), already performed three flybys on January 14, October 6, 2008, and September 29, 2009, before orbit insertion in March 2011. The second one, BepiColombo (ESA/JAXA), is planned to be launched in 2014 and to reach Mercury in 2020. The preparation of these two missions motivated an in-depth study of the rotation of Mercury.

\par The rotation of Mercury is a unique case in the Solar System because of its $3:2$ spin-orbit resonance, Mercury performing exactly 3 rotations during 2 revolutions about the Sun \citep{pd65}. It corresponds to an equilibrium state \citep{c65} known as Cassini State 1. Recently, radar Earth-based measurements by \citet{mpjsh07} detected a 88-day longitudinal libration of Mercury with an amplitude $\phi$ of $35.8\pm2$ arcsec. This amplitude being nearly twice too high to be consistent with a rigid Mercury, it is the signature of an at least partially molten core. If we consider Mercury as a 2-layered body with a rigid mantle and a spherical liquid core that does not follow the short-period ($\approx88$ days) excitations and does not interact with the mantle, we can derive from this amplitude the inertia of the mantle plus crust. In particular, naming $C_m$ the inertial polar momentum of the mantle and $A<B<C$ the inertial momenta of Mercury, we have \citep{p72}: 

\begin{equation}
\phi \approx 6\,C_{22}\frac{MR^2}{C}\frac{C}{C_m}\Big(1-11e^2+\frac{959}{48}e^4\Big)=\frac{3}{2}\frac{B-A}{C_m}\Big(1-11e^2+\frac{959}{48}e^4\Big)
\label{equ:peale72}
\end{equation}
where $C_{22}$ is a second-degree coefficient of the gravitational potential of Mercury, $M$ its mass, $R$ its radius, and $e$ its orbital eccentricity. It leads to $(B-A)/C_m \approx (2.033\pm0.114)\times10^{-4}$ with $e\approx 0.206$. If we take $C/MR^2=0.34$ \citep{mrvvb01} and $C_{22}=(1.0\pm0.5)\times10^{-5}$ \citep{acelt87} we get $C_m/C=0.579^{+0.339}_{-0.305}$.

\par Recent studies in one \citep{pmy09} and two \citep{dnrl09} degrees of freedom have theoretically estimated the longitudinal librations of Mercury. They highlighted in particular the possibility of a resonance with the jovian perturbation, whose period is $11.86$ years, that could potentially raise the amplitude of a long-term ($\approx12$ years) libration. Other periodic terms of a few arcsec have been estimated. This model also predicts that the latitudinal motion of Mercury should adiabatically follow the Cassini State 1 (\citet{p06}, \citet{dl08}), with short-period librations of about 10 milli-arcsec \citep{dnrl09}. In all these studies, the core-mantle interactions are neglected.

\par Recently, \citet{rvdb07} explored the dynamics of the rotation of Mercury, including core-mantle interactions in the SONyR model \citep{rb04}. We here propose an alternative study, starting from the Hamiltonian formulation of \citet{tw01} and highlighting the dynamical implications of core-mantle interactions, by considering Mercury as composed of a rigid mantle and a triaxial ellipsoidal cavity filled with inviscid fluid of constant uniform density and vorticity.

\section{The interior model}

\par The differential equations ruling the motion of a 2-layered body with a rigid mantle and a liquid non-spherical core have been derived by \citet{h95} and \citet{p10}. More recently, \citet{tw01} gave a Hamiltonian formulation of this problem, that \citet{h08} applied to the rotational dynamics of Io, assuming that the core and the mantle were aligned and proportional. Here, we generalize the model of Henrard, allowing the core to be non-proportional and non-spherical.

\subsection{Physical model}

\begin{figure}
\centering
\begin{tabular}{cc}
\includegraphics[height=5cm,width=7.8cm]{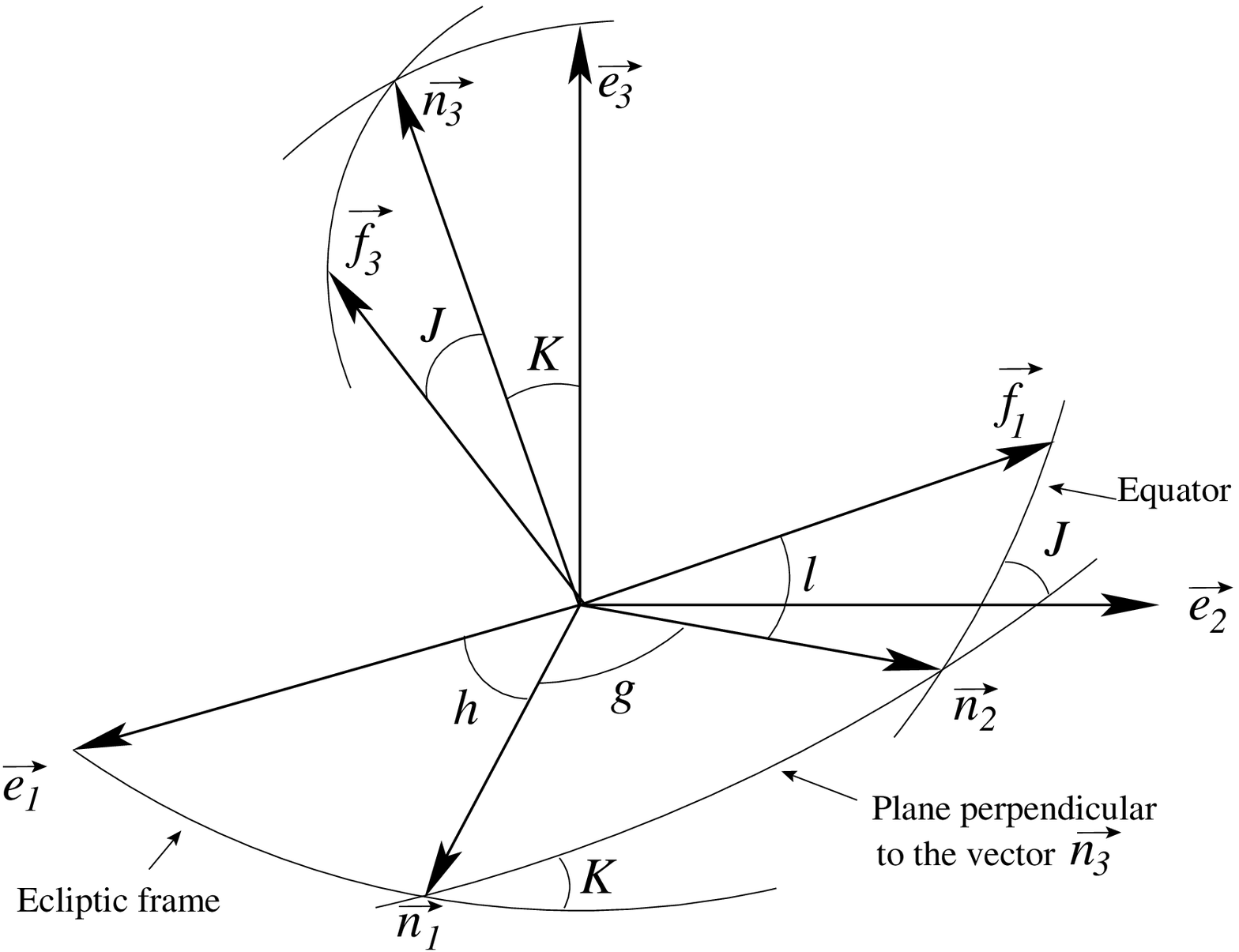} & \includegraphics[height=5cm,width=7.8cm]{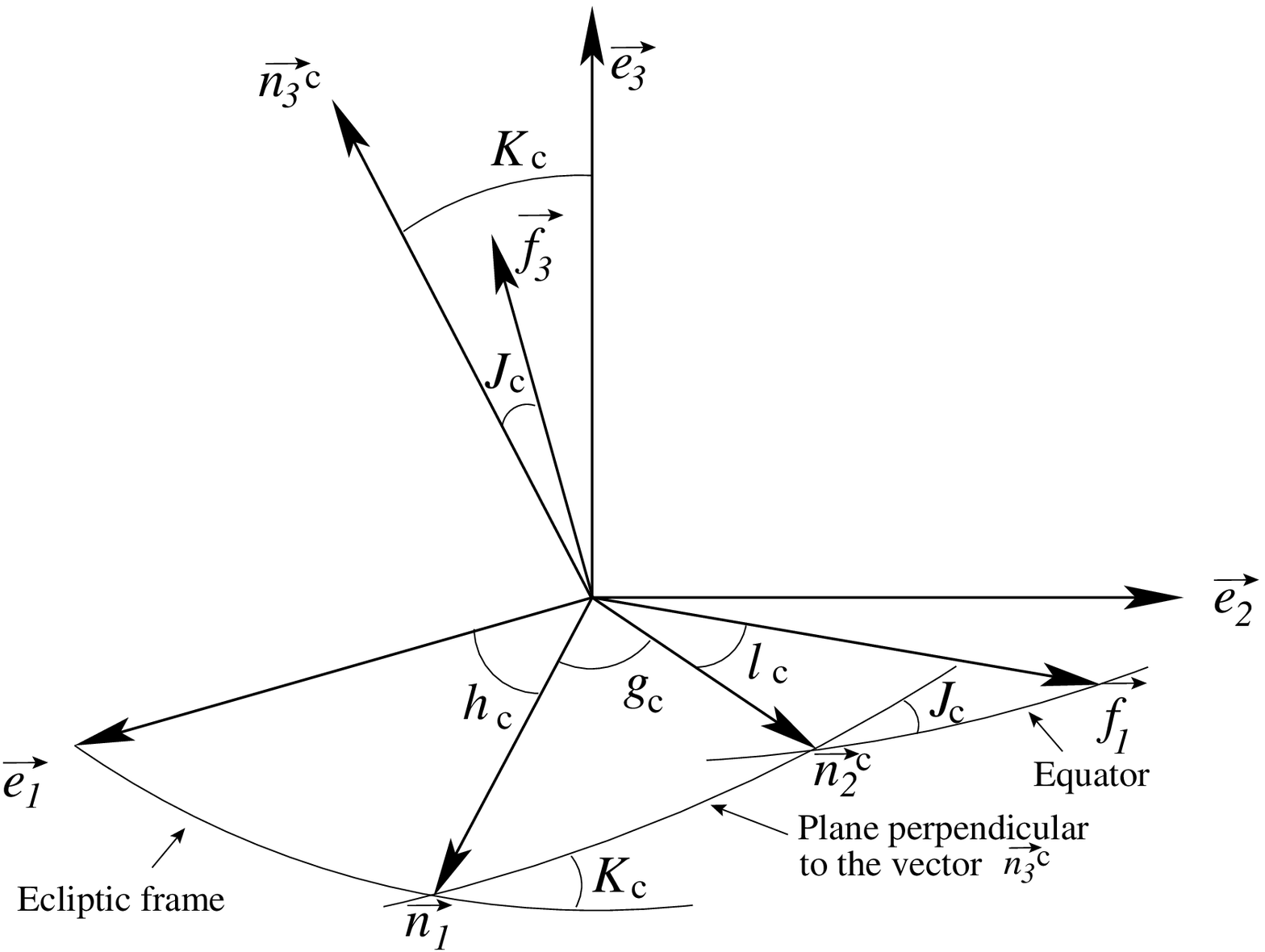}
\end{tabular}
\caption{In the left panel we have 3 reference frames: one linked to the ecliptic plane ($\vec{e_1},\vec{e_2},\vec{e_3}$), another linked to the angular momentum $\vec N$ ($\vec{n_1},\vec{n_2},\vec{n_3}$), and the last one linked to Mercury's axes of inertia ($\vec{f_1},\vec{f_2},\vec{f_3}$). In the right panel we have a similar configuration but instead of the angular momentum of Mercury, we have a reference frame linked to the angular momentum of a pseudo-core (defined later). We have the Euler angles $(h,K,g)$ positioning the vector $\vec{n_2}$ on the plane perpendicular to the angular momentum of Mercury and the Euler angles $(h_c,K_c,g_c)$ positioning the vector $\vec{n_2^c}$ on the plane perpendicular to the angular momentum of the pseudo-core. The angles $(l,J)$ and $(l_c,J_c)$ position the axis of least inertia. Note that $J_c$ is defined on the other side than $J$.   \label{fig:rot}}
\end{figure}

\par Four references frames are being considered (see Fig.\ref{fig:rot} \& \ref{fig:bigfig}). The first one, $(\vec{e_1},\vec{e_2},\vec{e_3})$ is assumed to be inertial for the rotational dynamics, it is in fact centered on Mercury and in translation with the inertial reference frame in which the orbital ephemerides of Mercury are given. This reference frame is related to the ecliptic at J2000. The second one, $(\vec{n^c_1},\vec{n^c_2},\vec{n^c_3})$ is linked to the angular momentum of a \emph{pseudo-core} that we define later, while the third one, i.e. $(\vec{n_1},\vec{n_2},\vec{n_3})$, is linked to the total angular momentum of Mercury. Finally, the last one, written as $(\vec{f_1},\vec{f_2},\vec{f_3})$, is rigidly linked to the principal axes of inertia of Mercury. In this last reference frame, the matrix of inertia of Mercury reads:

\begin{equation}
I=\left(\begin{array}{ccc}
A & 0 & 0 \\
0 & B & 0 \\
0 & 0 & C
\end{array}\right)
\label{equ:inertim}
\end{equation}
with $0<A\leq B \leq C$, while this of the core is:

\begin{equation}
I_c=\left(\begin{array}{ccc}
A_c & 0 & 0 \\
0 & B_c & 0 \\
0 & 0 & C_c
\end{array}\right),
\label{equ:inertic}
\end{equation}
in the same reference frame. So, the orientations of the mantle and the cavity are the same, a misalignment of their principal axes would require to consider the mantle as elastic, this is beyond the scope of the paper. As for the whole Mercury, we have $0<A_c\leq B_c \leq C_c$. In this way, the principal moments of inertia of the mantle are respectively $A_m=A-A_c$, $B_m=B-B_c$ and $C_m=C-C_c$ The principal elliptical radii of the cavity are written respectively $a$, $b$, $c$, yielding

\begin{center}
$\begin{array}{ccccc}
A_c & = & \displaystyle\iiint(x_2^2+x_3^2)\rho\,dx_1\,dx_2\,dx_3 & = & \frac{M_c}{5}(b^2+c^2), \label{equ:Ac} \\
B_c & = & \displaystyle\iiint(x_1^2+x_3^2)\rho\,dx_1\,dx_2\,dx_3 & = &  \frac{M_c}{5}(a^2+c^2), \label{equ:Bc} \\
C_c & = & \displaystyle\iiint(x_1^2+x_2^2)\rho\,dx_1\,dx_2\,dx_3 & = &  \frac{M_c}{5}(a^2+b^2), \label{equ:Cc}
\end{array}$
\end{center}
where $\rho$ is the density of mass of the fluid core, the integration being performed over the volume of the core.

\subsection{The kinetic energy of the system}

\par A Hamiltonian formulation of such a problem is usually composed of a kinetic energy and a disturbing potential, here the solar perturbation. Therefore, we consider every internal process, as the core-mantle interactions in our case, as part of the kinetic energy of Mercury. This section is widely inspired from \citet{h08}.

\par The components $(v_1,v_2,v_3)$ of the velocity field at the location $x_i$ inside the liquid core, in the frame of the principal axes of inertia of the mantle, are assumed to be:

\begin{eqnarray}
v_1 & = & \Big(\omega_2+\frac{a}{c}\nu_2\Big)x_3-\Big(\omega_3+\frac{a}{b}\nu_3\Big)x_2, \label{equ:v1} \\
v_2 & = & \Big(\omega_3+\frac{b}{a}\nu_3\Big)x_1-\Big(\omega_1+\frac{b}{c}\nu_1\Big)x_3, \label{equ:v2} \\
v_3 & = & \Big(\omega_1+\frac{c}{b}\nu_1\Big)x_2-\Big(\omega_2+\frac{c}{a}\nu_2\Big)x_1, \label{equ:v3}
\end{eqnarray}
where $(\omega_1,\omega_2,\omega_3)$ are the components of the angular velocity of the mantle with respect to an inertial frame, and the vector of coordinates $(\nu_1,\nu_2,\nu_3)$ specifies the velocity field of the core with respect to the moving mantle.

\par  The angular momentum of the core $\vec{N'_c}$ is obtained by:

\begin{equation}
\vec{N'_c}=\iiint_{core}(\vec{x}\times\vec{v})\rho\,dx_1\,dx_2\,dx_3
\label{equ:intNpc}
\end{equation}
and the result is:

\begin{equation}
\begin{split}
\vec{N'_c}= \frac{M_c}{5}\Bigg[\bigg(\frac{c}{b}\nu_1+\omega_1\bigg)b^2+\bigg(\frac{b}{c}\nu_1+\omega_1\bigg)c^2\Bigg]\vec{f_1} \\
 +\frac{M_c}{5}\Bigg[\bigg(\frac{c}{a}\nu_2+\omega_2\bigg)a^2+\bigg(\frac{a}{c}\nu_2+\omega_2\bigg)c^2\Bigg]\vec{f_2} \\
 +\frac{M_c}{5}\Bigg[\bigg(\frac{b}{a}\nu_3+\omega_3\bigg)a^2+\bigg(\frac{a}{b}\nu_3+\omega_3\bigg)b^2\Bigg]\vec{f_3}. \label{equ:Npc}
\end{split}
\end{equation}
We now set the following quantities:

\begin{center}
$\begin{array}{ccccc}
D_1 &=& \frac{2M_c}{5}bc &=& \sqrt{\big(A_c-B_c+C_c\big)\big(A_c+B_c-C_c\big)} \\
D_2 &=& \frac{2M_c}{5}ac &=& \sqrt{\big(-A_c+B_c+C_c\big)\big(A_c+B_c-C_c\big)} \\
D_3 &=& \frac{2M_c}{5}ab &=& \sqrt{\big(-A_c+B_c+C_c\big)\big(A_c-B_c+C_c\big)}
\end{array}$
\end{center}
and we can write:

\begin{equation}
\vec{N'_c}=\big[A_c\omega_1+D_1\nu_1\big]\vec{f_1}+\big[B_c\omega_2+D_2\nu_2\big]\vec{f_2}+\big[C_c\omega_3+D_3\nu_3\big]\vec{f_3},
\label{equ:NPc2}
\end{equation}
while the angular momentum of the mantle is

\begin{equation}
\vec{N_m}=A_m\omega_1\vec{f_1}+B_m\omega_2\vec{f_2}+C_m\omega_3\vec{f_3},
\label{equ:Nman}
\end{equation}
and the total angular momentum of Mercury is

\begin{equation}
\vec{N}=\big[A\omega_1+D_1\nu_1\big]\vec{f_1}+\big[B\omega_2+D_2\nu_2\big]\vec{f_2}+\big[C\omega_3+D_3\nu_3\big]\vec{f_3}.
\label{equ:Ntot}
\end{equation}

\par The kinetic energy of the core is

\begin{equation}
T_c=\frac{1}{2}\iiint_{core}\rho v^2\,dx_1\,dx_2\,dx_3
\label{equ:tc}
\end{equation}
i.e.

\begin{equation}
T_c=\frac{1}{2}\Big(A_c(\omega_1^2+\nu_1^2)+B_c(\omega_2^2+\nu_2^2)+C_c(\omega_3^2+\nu_3^2)+D_1\omega_1\nu_1+D_2\omega_2\nu_2+D_3\omega_3\nu_3\Big),
\label{equ:tc2}
\end{equation}
while the kinetic energy of the mantle $T_m$ is

\begin{equation}
T_m=\frac{1}{2}\vec{N_m}\cdot\vec{\omega}=\frac{A_m\omega_1^2+B_m\omega_2^2+C_m\omega_3^2}{2}.
\label{equ:Tm}
\end{equation}
From $T=T_m+T_c$ we finally deduce the kinetic energy of Mercury:

\begin{equation}
T=\frac{1}{2}\big(A\omega_1^2+B\omega_2^2+C\omega_3^2+A_c\nu_1^2+B_c\nu_2^2+C_c\nu_3^2+2D_1\omega_1\nu_1+2D_2\omega_2\nu_2+2D_3\omega_3\nu_3\big).
\label{equ:T}
\end{equation}

\par We can easily check the expressions of the partial derivatives, as

\begin{equation}
\frac{\partial T}{\partial \omega_1}  =  A\omega_1+D_1\nu_1  =  N_1
\label{equ:N1}
\end{equation}
or 

\begin{equation}
\frac{\partial T}{\partial \nu_1}  =  D_1\omega_1+A_c\nu_1  =  N_1^c,
\label{equ:N1c}
\end{equation}
where $N_i$ are the components of the total angular momentum. $N_i^c$ are not the components of the angular momentum of the core but are close to it for a cavity close to spherical. We have, for instance for the first component:

\begin{equation}
N_1^c-N_1'^c=(A_c-D_1)(\omega_1-\nu_1)=\frac{M_c}{5}(c-b)^2(\omega_1-\nu_1),
\label{equ:depart}
\end{equation}
so the difference is of the second order in departure from the sphericity. From now on, we call \textit{angular momentum of the pseudo-core} the vector $\vec{N^c}=N_1^c\vec{f_1}+N_2^c\vec{f_2}+N_3^c\vec{f_3}$.

\par With these notations, the Poincar\'e-Hough's equations of motion, for the system mantle-core in the absence of external torque, are (see e.g. Eq.15 in \citet{tw01} or \citet{h08}):

\begin{eqnarray}
\frac{d\vec{N}}{dt} & = & \vec{N} \times \vec{\nabla}_{\vec{N}}\mathcal{T}, \label{equ:ph1} \\
\frac{d\vec{N_c}}{dt} & = & \vec{N_c} \times \vec{\nabla}_{-\vec{N_c}}\mathcal{T}, \label{equ:ph2}
\end{eqnarray}
with
\begin{equation}
\label{equ:gradient}
\vec{\nabla}_{\vec{N}}\mathcal{T}=\frac{\partial \mathcal{T}}{\partial N_1}\vec{f_1}+\frac{\partial \mathcal{T}}{\partial N_2}\vec{f_2}+\frac{\partial \mathcal{T}}{\partial N_3}\vec{f_3}.
\end{equation}
Here $\mathcal{T}$ is the kinetic energy expressed in terms of the components of the vectors $\vec{N}$ and $\vec{N_c}$, i.e.

\begin{equation}
\begin{split}
\mathcal{T}=\frac{1}{2\alpha}\big(A_cN_1^2+A(N_1^c)^2-2D_1N_1N_1^c\big)+\frac{1}{2\beta}\big(B_cN_2^2+B(N_2^c)^2-2D_2N_2N_2^c\big) \\
+\frac{1}{2\gamma}\big(C_cN_3^2+C(N_3^c)^2-2D_3N_3N_3^c\big)
\end{split}
\end{equation}
with $\alpha=AA_c-D_1^2$, $\beta=BB_c-D_2^2$ and $\gamma=CC_c-D_3^2$.

\subsection{The Hamiltonian}

\subsubsection{The rotational kinetic energy}

\par We assume that the cavity and Mercury are almost spherical, this allows us to introduce the four small parameters $\epsilon_i$:

\begin{eqnarray}
\epsilon_1 &=& \frac{2C-A-B}{2C}=J_2\frac{MR^2}{C}, \label{equ:eps1} \\
\epsilon_2 &=& \frac{B-A}{2C}=2C_{22}\frac{MR^2}{C}, \label{equ:eps2} \\
\epsilon_3 &=& \frac{2C_c-A_c-B_c}{2C_c}, \label{equ:eps3} \\
\epsilon_4 &=& \frac{B_c-A_c}{2C_c}, \label{equ:eps4}
\end{eqnarray}
and also the parameter $\delta=C_c/C$, i.e. the ratio between the polar inertial momentum of the core and of Mercury. $\epsilon_1$ represents the polar flattening of Mercury, while $\epsilon_2$ is its equatorial ellipticity. $\epsilon_3$ and $\epsilon_4$ have the same meaning for the cavity. If we assume the core of Mercury to be spherical, we should take $\epsilon_3=\epsilon_4=0$, while $\epsilon_4=0$ represents an axisymmetric cavity. \citet{h08} considered that the ellipsoid of inertia of the core and the mantle were aligned and proportional, the mathematical formulation was $\epsilon_3=\epsilon_1$ and $\epsilon_4=\epsilon_2$. Our parameters are gathered in Table \ref{tab:values}.

\begin{table}
\centering
\caption{The shape parameters of Mercury. \label{tab:values}}
\begin{tabular}{lll}
\hline\hline
Parameter & Value & Reference \\
\hline
$J_2$ & $(6.0\pm2.0)\times10^{-5}$ & \citet{acelt87} \\
$C_{22}$ & $(1.0\pm0.5)\times10^{-5}$ & \citet{acelt87} \\
$C/(MR^2)$ & $0.34$ & \citet{mrvvb01} \\
$\delta=1-C_m/C$ & $0.421$ & \citet{mpjsh07} \\
$\epsilon_1=J_2MR^2/C$ & $1.765\times10^{-4}$ & -- \\
$\epsilon_2=2C_{22}MR^2/C$ & $5.882\times10^{-5}$ & -- \\
\hline
\end{tabular}
\end{table}

\par
\begin{figure}
\centering
\includegraphics[width=10cm]{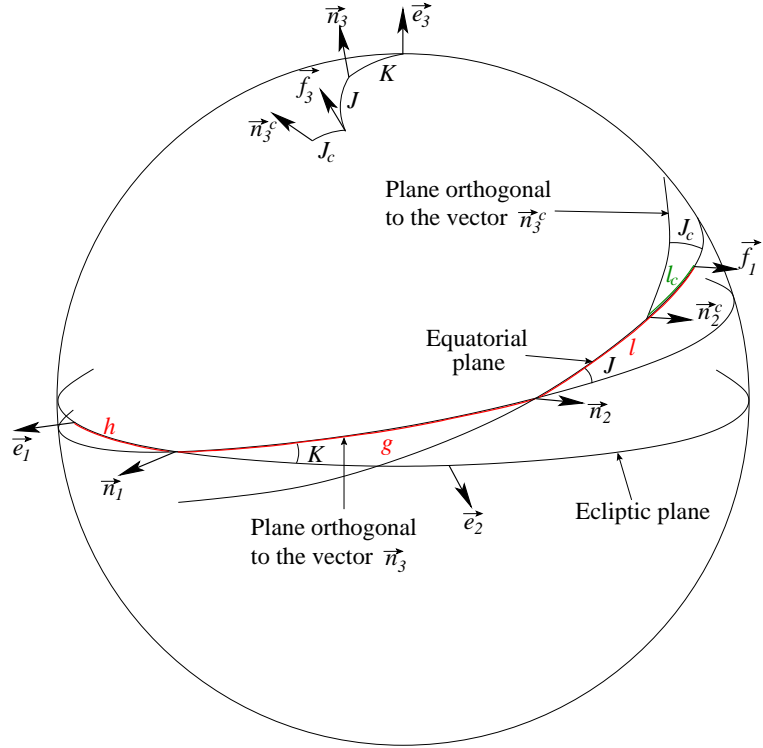}
\caption{The four reference frames gathered in the same view. The angles $(h,K)$ position the plane orthogonal to the angular momentum $\vec N$. The Euler angles $(g,J,l)$ locate the axis of least inertia and the body frame $(\vec{f_1},\vec{f_2},\vec{f_3})$. The angles $(J_c,l_c)$ place the angular momentum of the pseudo-core with respect to the axis of least inertia $f_1$.\label{fig:bigfig}}
\end{figure}

 We now introduce the two sets of Andoyer's variables \citep{a26}, $(l,g,h,L,G,H)$ and $(l_c,g_c,h_c,L_c,G_c,H_c)$, related respectively to the whole Mercury and to its core. The angles $(h,K,g)$ are the Euler angles of the vector $\vec{n_2}$, node of the equatorial plane over the plane perpendicular to the angular momentum $\vec{N}$, the angles $(J,l)$ position the axis of least inertia $\vec{f_1}$ with respect to $\vec{n_2}$. Correspondingly the angles $(h_c,K_c,g_c)$ are the Euler angles of the vector $\vec{n^c_2}$, node of the equatorial plane over the plane perpendicular to the angular momentum of the pseudo-core $\vec{N_c}$, and $(J_c,l_c)$ position the axis of least inertia with respect to $\vec{n^c_2}$. Figure \ref{fig:bigfig} shows a schematic view of all the reference frames and relevant angles. The variables are $(h,g,l)$ and $(h_c,g_c,l_c)$ and the corresponding momenta ($H=N\cos K$, $G=N$, $L=N\cos J$) and ($H_c=N^c \cos K_c$, $G_c=N^c$, $L_c=N^c \cos J_c$). Expressed in Andoyer's variables the components of $\vec{N}$ and $\vec{N^c}$ are:

\begin{center}
$\begin{array}{lll}
N_1=\sqrt{G^2-L^2}\sin l, & \hspace{2cm} & N_1^c=\sqrt{G_c^2-L_c^2}\sin l_c, \\
N_2=\sqrt{G^2-L^2}\cos l, & \hspace{2cm} & N_2^c=\sqrt{G_c^2-L_c^2}\cos l_c, \\
N_3=L, & \hspace{2cm} & N_3^c=L_c. \\
\end{array}$
\end{center}

\par We can now straightforwardly derive the Hamiltonian $\mathcal{H}_1$ of the free rotation of Mercury, using Andoyer's variables and changing the sign of $\vec{N^c}$ to take the minus sign of the Poincar\'e-Hough equations into account (Eq.\ref{equ:ph2}). We also linearize the Hamiltonian with respect to the small parameters $\epsilon_i$ (their orders of magnitude being about $10^{-5}$), and get:

\begin{eqnarray}
\mathcal{H}_0&=&\frac{1}{2C(1-\delta)}\bigg(G^2+\frac{G_c^2}{\delta}+2\sqrt{(G^2-L^2)(G_c^2-L_c^2)}\cos(l-l_c)+2LL_c\bigg) \nonumber \\
&+&\frac{\epsilon_1}{2C(1-\delta)^2}\bigg(G^2-L^2+G_c^2-L_c^2+2\sqrt{(G^2-L^2)(G_c^2-L_c^2)}\cos(l-l_c)\bigg) \nonumber \\
&-&\frac{\epsilon_2}{2C(1-\delta)^2}\bigg((G^2-L^2)\cos(2l)+(G_c^2-L_c^2)\cos(2l_c)+2\sqrt{(G^2-L^2)(G_c^2-L_c^2)}\cos(l+l_c)\bigg) \nonumber \\
&-&\frac{\epsilon_3}{2C(1-\delta)^2}\bigg(\delta (G^2-L^2)+(G_c^2-L_c^2)(2-\frac{1}{\delta})+2\delta\sqrt{(G^2-L^2)(G_c^2-L_c^2)}\cos(l-l_c)\bigg) \nonumber \\
&+&\frac{\epsilon_4}{2C(1-\delta)^2}\bigg(\delta(G^2-L^2)\cos(2l)+(G_c^2-L_c^2)(2-\frac{1}{\delta})\cos(2l_c) \nonumber \\
& & +2\delta\sqrt{(G^2-L^2)(G_c^2-L_c^2)}\cos(l+l_c)\bigg). \label{equ:HG}
\end{eqnarray}

\par We now introduce the following canonical change of variables, of multiplier $\frac{1}{nC}$, $n$ being the mean orbital motion of Mercury:
\begin{equation}\label{eq:chvar}
\begin{array}{lll}
p=l+g+h, & \hspace{2cm} & P=\frac{G}{nC}, \\
r=-h, & \hspace{2cm} & R=P(1-\cos K), \\
\xi_1=-\sqrt{2P(1-\cos J)}\sin l, & \hspace{2cm} & \eta_1=\sqrt{2P(1-\cos J)}\cos l, \\
p_c=-l_c+g_c+h_c, & \hspace{2cm} & P_c=\frac{G_c}{nC}, \\
r_c=-h_c, & \hspace{2cm} & R_c=P_c(1-\cos K_c), \\
\xi_2=\sqrt{2P_c(1+\cos J_c)}\sin l_c, & \hspace{2cm} & \eta_2=\sqrt{2P_c(1+\cos J_c)}\cos l_c. \\
\end{array} \\
\end{equation}

\par In order to be consistent with the sign minus in the equations and before $l_c$, the wobble of the pseudo-core $J_c$ has to be replaced by $\pi-J_c$. In this way, we have $L_c=G_c\cos(\pi-J_c)=-G_c\cos(J_c)$. In this new set of variables, we have

\begin{center}
$\begin{array}{lll}
N_1=-nC\sqrt{P^2-\Big(P-\frac{\xi_1^2+\eta_1^2}{2}\Big)^2}\frac{\xi_1}{\xi_1^2+\eta_1^2}, & \hspace{1.5cm} & N_1^c=nC\sqrt{P_c^2-\Big(\frac{\xi_2^2+\eta_2^2}{2}-P_c\Big)^2}\frac{\xi_2}{\xi_2^2+\eta_2^2}, \\
N_2=nC\sqrt{P^2-\Big(P-\frac{\xi_1^2+\eta_1^2}{2}\Big)^2}\frac{\eta_1}{\xi_1^2+\eta_1^2}, & \hspace{1.5cm} & N_2^c=nC\sqrt{P_c^2-\Big(\frac{\xi_2^2+\eta_2^2}{2}-P_c\Big)^2}\frac{\eta_2}{\xi_2^2+\eta_2^2}, \\
N_3=nC\Big(P-\frac{\xi_1^2+\eta_1^2}{2}\Big), & \hspace{1.5cm} & N_3^c=nC\Big(\frac{\xi_2^2+\eta_2^2}{2}-Pc\Big), \\
\end{array}$ \\
\end{center}
and the Hamiltonian of the free rotational motion becomes, after division by $nC$: 
\begin{eqnarray}
\mathcal{H}_1&=&\frac{n}{2(1-\delta)}\Bigg(P^2+\frac{P_c^2}{\delta}+2\sqrt{\Big(P-\frac{\xi_1^2+\eta_1^2}{4}\Big)\Big(P_c-\frac{\xi_2^2+\eta_2^2}{4}\Big)}\big(\eta_1\eta_2-\xi_1\xi_2\big) \nonumber \\
 & & +2\Big(P-\frac{\xi_1^2+\eta_1^2}{2}\Big)\Big(\frac{\xi_2^2+\eta_2^2}{2}-P_c\Big)\Bigg) \nonumber \\
%& & \nonumber \\
&+&\frac{n\epsilon_1}{2(1-\delta)^2}\Bigg(P_c^2-\Big(\frac{\xi_2^2+\eta_2^2}{2}-P_c\Big)^2+P^2-\Big(P-\frac{\xi_1^2+\eta_1^2}{2}\Big)^2 \nonumber \\
&+& 2\sqrt{\Big(P-\frac{\xi_1^2+\eta_1^2}{4}\Big)\Big(P_c-\frac{\xi_2^2+\eta_2^2}{4}\Big)}\big(\eta_1\eta_2-\xi_1\xi_2\big)\Bigg) \nonumber \\
%& & \nonumber \\
&+&\frac{n\epsilon_2}{2(1-\delta)^2}\Bigg(\frac{1}{4}\big(4P-\xi_1^2-\eta_1^2\big)\big(\xi_1^2-\eta_1^2\big)+\frac{1}{4}\big(4P_c-\xi_2^2-\eta_2^2\big)\big(\xi_2^2-\eta_2^2\big) \nonumber \\
&-&2\sqrt{\Big(P-\frac{\xi_1^2+\eta_1^2}{4}\Big)\Big(P_c-\frac{\xi_2^2+\eta_2^2}{4}\Big)}\big(\eta_1\eta_2+\xi_1\xi_2\big)\Bigg) \nonumber \\
%& & \nonumber \\
&-&\frac{n\epsilon_3}{2(1-\delta)^2}\Bigg(\delta \Big(P^2-\Big(P-\frac{\xi_1^2+\eta_1^2}{2}\Big)^2\Big)+\Big(P_c^2-(\frac{\xi_2^2+\eta_2^2}{2}-P_c\Big)^2\Big)\Big(2-\frac{1}{\delta}\Big) \nonumber \\
&+& 2\delta\sqrt{\Big(P-\frac{\xi_1^2+\eta_1^2}{4}\Big)\Big(P_c-\frac{\xi_2^2+\eta_2^2}{4}\Big)}\big(\eta_1\eta_2-\xi_1\xi_2\big)\Bigg) \nonumber \\
%& & \nonumber \\
&+&\frac{n\epsilon_4}{2(1-\delta)^2}\Bigg(\frac{\delta}{4}\big(4P-\xi_1^2-\eta_1^2\big)\big(\eta_1^2-\xi_1^2\big)+\Big(2-\frac{1}{\delta}\Big) \frac{1}{4}\big(4P_c-\xi_2^2-\eta_2^2\big)\big(\eta_2^2-\xi_2^2\big)  \nonumber \\
&+ & 2\delta\sqrt{\Big(P-\frac{\xi_1^2+\eta_1^2}{4}\Big)\Big(P_c-\frac{\xi_2^2+\eta_2^2}{4}\Big)}\big(\eta_1\eta_2+\xi_1\xi_2\big)\Bigg). \label{equ:HG3}
\end{eqnarray}

\par Finally, in order to get an easy-to-use formula, we can develop this Hamiltonian up to the second order in ($\xi_1$, $\xi_2$, $\eta_1$, $\eta_2$) to get:

\begin{eqnarray}
\mathcal{H}_1&=&\frac{n}{2(1-\delta)}\Bigg(P^2+\frac{P_c^2}{\delta}+2\sqrt{PP_c}\big(\eta_1\eta_2-\xi_1\xi_2\big)+2\Big(P\frac{\xi_2^2+\eta_2^2}{2}+P_c\frac{\xi_1^2+\eta_1^2}{2}-PP_c\Big)\Bigg) \nonumber \\
& & \nonumber \\
&+&\frac{n\epsilon_1}{2(1-\delta)^2}\Bigg(P\big(\xi_1^2+\eta_1^2\big)+P_c\big(\xi_2^2+\eta_2^2\big)+2\sqrt{PP_c}\big(\eta_1\eta_2-\xi_1\xi_2\big)\Bigg) \nonumber \\
& & \nonumber \\
&+&\frac{n\epsilon_2}{2(1-\delta)^2}\Bigg(P\big(\xi_1^2-\eta_1^2\big)+P_c\big(\xi_2^2-\eta_2^2\big)-2\sqrt{PP_c}\big(\eta_1\eta_2+\xi_1\xi_2\big)\Bigg) \nonumber \\
& & \nonumber \\
&-&\frac{n\epsilon_3}{2(1-\delta)^2}\Bigg(\delta P\big(\xi_1^2+\eta_1^2\big)+\Big(2-\frac{1}{\delta}\Big)P_c \big(\xi_2^2+\eta_2^2\big)+2\delta\sqrt{PP_c}\big(\eta_1\eta_2-\xi_1\xi_2\big)\Bigg) \nonumber \\
& & \nonumber \\
&+&\frac{n\epsilon_4}{2(1-\delta)^2}\Bigg(\delta P\big(\eta_1^2-\xi_1^2\big)+\Big(2-\frac{1}{\delta}\Big)P_c\big(\eta_2^2-\xi_2^2\big)+2\delta\sqrt{PP_c}\big(\eta_1\eta_2+\xi_1\xi_2\big)\Bigg). \label{equ:HG4}
\end{eqnarray}

\subsubsection{The gravitational potential}

To compute the gravitational potential due to the Solar perturbation on Mercury we must first obtain the coordinates $x$, $y$, and $z$ of the Sun in the reference frame linked to the principal axes of inertia $(\vec{f_1},\vec{f_2},\vec{f_3})$. Five rotations are to be performed:

\begin{equation}
\left(\begin{array}{c}
x \\
y \\
z
\end{array}\right)
=R_3(-l)R_1(-J)R_3(-g)R_1(-K)R_3(-h)\left(\begin{array}{c}
x_i \\
y_i \\
z_i
\end{array}\right)
\label{equ:passage}
\end{equation}
with $x_i$, $y_i$, $z_i$ depending on the mean anomaly $l_o$, the longitude of the ascending node $\ascnode_o$, the longitude of the perihelion $\varpi_o$, the inclination $i$, and the eccentricity $e$.\\
The rotation matrices are defined by

\begin{equation}
R_3(\phi)=\left(\begin{array}{ccc}
\cos\phi & -\sin\phi & 0 \\
\sin\phi & \cos\phi & 0 \\
0 & 0 & 1
\end{array}\right),\qquad
R_1(\phi)=\left(\begin{array}{ccc}
1 & 0 & 0 \\
0 & \cos\phi & -\sin\phi \\
0 & \sin\phi & \cos\phi 
\end{array}\right).
\label{equ:r3}
\end{equation}
The gravitational potential then reads:
\begin{equation}
V_1(l_o,l,g,h,J,K)=-\frac{3}{2}C\frac{\mathcal{G}M}{d^3}\big(\epsilon_1(x^2+y^2)+\epsilon_2(x^2-y^2)\big)
\label{equ:pull1}
\end{equation}
where $\mathcal{G}$ is the gravitational constant, $M$ the mass of the Sun, $(x,y,z)$ the unit vector pointing at the Sun in the frame $(\vec{f_1},\vec{f_2},\vec{f_3})$, while $d$ is the distance Sun-Mercury (expanded in eccentricity and mean anomaly). \\
Let us note that unlike \citet{h08}, we consider that the perturbation is applied to the whole planet and not only to its mantle. We address the dynamical consequences later in the paper.\\
From the variables $x$, $y$ and $z$, it is easy to introduce the set of variables defined in (\ref{eq:chvar}). We also modify the moment $\Lambda_o$ associated with $l_o$ (that appears in the expressions of $x$ and $y$) in such way that all our variables are now canonical with multiplier $1/nC$ and our gravitational potential becomes (after division by $nC$)
\begin{equation}
\mathcal{H}_2(l_o,p,r,R,\xi_1,\eta_1)=-\frac{3}{2}\frac{\mathcal{G}M}{nd^3}\big(\epsilon_1(x^2+y^2)+\epsilon_2(x^2-y^2)\big).
\label{equ:pull2}
\end{equation}

\par Finally, we use the formulae (\ref{equ:HG4}) and (\ref{equ:pull2}) to get the Hamiltonian of the system:

\begin{equation}
\mathcal{H}=\mathcal{H}_1(P,\xi_1,\eta_1,\xi_2,\eta_2)+\mathcal{H}_2(l_o,p,r,R,\xi_1,\eta_1).
\label{equ:hamiltout}
\end{equation}
The four degrees of freedom of this Hamiltonian are the spin ($p$, $P$), the obliquity ($r$, $R$), the wobble of the whole body ($\xi_1$, $\eta_1$) and the wobble of the core ($\xi_2$, $\eta_2$). In this study, we name "wobble" every motion dealing with a shift between the angular momentum of the body or its core, and its geometrical pole axis. It is different from the polar motion that concerns the rotation axis instead of the angular momentum. Contrary to the Chandler wobble for the Earth, we include in the term "wobble" every periodic contribution constituting this motion.

\section{Comparison between an analytical and a numerical study}

\par To study this problem, we use both analytical and numerical methods that allow us to compare their efficiencies and check the reliability of the results.

\subsection{Analytical study}

\par In a previous paper by the authors \citep{dnrl09}, our model was a 2-degree of freedom Hamiltonian neglecting the wobble $J$, but including the planetary perturbations. Here we have a 4-degree of freedom Hamiltonian, but the way we perform our analytical study is similar to our previous paper. However there are some key differences that we will highlight in this section. All the computations were made using our algebraic manipulator called MSNam \citep{h86}.

\subsubsection{Resonant angles and Hamiltonian}

\par As mentioned earlier, it is a known fact that Mercury is in a 3:2 spin-orbit resonance. In other words, the rotation speed of Mercury $\dot p$ (where $p=l+g+h$, the spin angle of Mercury) is 1.5 times larger than its mean motion  $n$ , i.e. $\dot p=\frac{3}{2}n$. The angle describing this resonance is $\sigma_1=l_o-\frac{3}{2}p-\varpi_o$, with $\varpi_o$ the longitude of the perihelion. The angle $\sigma_1$ actually represents the libration in longitude.\\
The second resonant angle characterizes the 1:1 commensurability between the orbital and rotational nodes, following the 3rd of Cassini's laws (\citet{c66} or \citet{ldr06} for Cassini's laws applied to Mercury): $\sigma_2=r+\Omega_o$, with $\Omega_o$ being the longitude of the ascending node. This angle is actually linked to the latitudinal motion of Mercury (through its conjugated moment $R$ which depends on the ecliptic obliquity $K$).\\
Introducing the resonant angles in the Hamiltonian (\ref{equ:hamiltout}) and using cartesian-like coordinates (expanded to order 5) for the 4 degrees of freedom:
\begin{equation}\label{eq:varcar}
\begin{array}{lll}
x_1=\sigma_1 \text{ (expanded around 0)}, & \hspace{2cm} & y_1=P, \\
x_2=\sqrt{2P(1-\cos K)}\sin \sigma_2, & \hspace{2cm} & y_2=\sqrt{2P(1-\cos K)}\cos \sigma_2, \\
\xi_1=-\sqrt{2P(1-\cos J)}\sin l, & \hspace{2cm} & \eta_1=\sqrt{2P(1-\cos J)}\cos l, \\
\xi_2=\sqrt{2P_c(1+\cos J_c)}\sin l_c, & \hspace{2cm} & \eta_2=\sqrt{2P_c(1+\cos J_c)}\cos l_c, 
\end{array} 
\end{equation}
the Hamiltonian is now $\mathcal H=\mathcal H(l_o,x_1,y_1,x_2,y_2,\xi_1,\eta_1,\xi_2,\eta_2)$.\\
We also add constant precessions of the perihelion and the node, respectively $d\varpi_o/dt=0.2772831860533198\times10^{-4}$ $rad/y$ and $d\ascnode_o/dt=-0.2189047296429404\times10^{-4}$ $rad/y$ from the VSOP planetary theory \citep{fs05}. These precessions, through the introduction of the resonant angle $\sigma_2$, will result in forced libration in latitude.

\subsubsection{Equilibria and free periods of the averaged quadratic Hamiltonian}
To compute the equilibria of the Hamiltonian, we first average the Hamiltonian over the fast angular variable (the mean anomaly $l_o$). Assuming that Mercury lies at the Cassini equilibrium and that there is no wobble motion (for the whole body and the core), we have $x_1=x_2=\xi_1=\eta_1=\xi_2=\eta_2=0$. Putting that into the Hamiltonian, we compute the equilibria of $y_1$ and $y_2$ using Hamilton's equations and an iterative process and we find $y_1^\star=1.5-6.117\times 10^{-7}$ and $y_2^\star=0.1502$, resulting in an ecliptic obliquity of $K^\star=7^\circ 1.873 \text{ arcmin}$.\\
After a translation to this equilibrium, our quadratic averaged Hamiltonian looks like this: 
\begin{equation}
\mathcal{\bar H}_2=ax_1^2+bx_1x_2+cx_2^2+dy_1+ey_1y_2+fy_2^2+g\xi_1^2+h\xi_1\xi_2+i\xi_2^2+j\eta_1+k\eta_1\eta_2+l\eta_2^2.
\end{equation}
We notice that the degrees of freedom related to the librations in longitude $(x_1,y_1)$ and latitude $(x_2,y_2)$, and those related to the wobbles of the planet $(\xi_1,\eta_1)$ and the core $(\xi_2,\eta_2)$ are coupled two by two, the coupling being much weaker in the first case than in the second one.\\
To compute the fundamental periods of this Hamiltonian, we must first disentangle the coupled degrees of freedom. To do this we use an untangling transformation \citep{hl05} twice, and after changes of variables to action-angle variables:
\begin{equation}
\left\{
\begin{array}{ll}
x_1=\sqrt{2U}\sin u, &y_1=\sqrt{2U}\cos u,\\
x_2=\sqrt{2V}\sin v, &y_2=\sqrt{2V}\cos v,\\
\xi_1=\sqrt{2W}\sin w, &\eta_1=\sqrt{2W}\cos w,\\
\xi_2=\sqrt{2Z}\sin z, &\eta_2=\sqrt{2Z}\cos z,
\end{array}\right.
\end{equation}
we have the following quadratic Hamiltonian:
\begin{equation}\label{h00}
\mathcal{\bar H}_2=n_u U+n_v V+n_w W+n_z Z,
\end{equation}
with $n_u$, $n_v$, $n_w$, $n_z$ the free frequencies corresponding respectively to the libration in longitude, the libration in latitude, the wobble of the planet and the wobble of the core.\\
These frequencies (especially $n_v$) actually depend on the shape of the core. For the parameters given in Table \ref{tab:values} and $\epsilon_3=\epsilon_1$ and $\epsilon_4=\epsilon_2$, the corresponding fundamental periods are
\begin{eqnarray}
&& T_u=12.0601\text{ years,} \qquad T_v=1065.99\text{ years,}\\
&& T_w=337.726\text{ years,} \qquad T_z=58.6219\text{ days.}
\end{eqnarray}

 \subsubsection{Lie perturbation theory}
To compute the evolution of the different variables, we use a perturbation theory by Lie transforms (see e.g. \citet{d69}). Our main Hamiltonian is the quadratic Hamiltonian described in the previous section and the perturbation contains all the other terms that we neglected to compute the fundamental frequencies (and to which we applied the same transformations). Here is a quick reminder on how this works.\\
This perturbation theory is visualised through a Lie triangle:
\begin{equation}
\begin{array}{cccl}
{\mathcal H}^0_0&&&\\
{\mathcal H}^0_1& {\mathcal H}^1_0& &\\
-& {\mathcal H}^1_1 & {\mathcal H}^2_0&\\
-& {\mathcal H}^1_2 & {\mathcal H}^2_1 & {\mathcal H}^3_0\\
\vdots& \vdots& \vdots &\;\;\vdots\quad \ddots
\end{array}
\end{equation}
In the part ${\mathcal H}^0_0$ we put the averaged quadratic Hamiltonian explained in the previous section. In ${\mathcal H}^0_1$ we put what remains, i.e. the perturbation containing the higher order terms in $U$, $V$, $W$, $Z$, and the short-period terms.\\
In the principal diagonal we will find the averaged Hamiltonian: $\bar {\mathcal H}=\sum_{i=0}^{\text{order}} {\mathcal H}^i_0/i!$. To get this averaged Hamiltonian we use the following homological equation:
\begin{eqnarray*}
{\mathcal H}^n_0&=&{\mathcal H}_1^{n-1}+\big({\mathcal H}_0^{n-1};W_1\big)
\end{eqnarray*}
with the intermediate ${\mathcal H}^n_j$ computed as follows:
\begin{eqnarray*}
{\mathcal H}_j^n&=&{\mathcal H}_{j+1}^{n-1}+\sum_{i=0}^{j}\left(\begin{array}{c}\!\!j\!\!\\\!\!i\!\!\end{array}\right)  \big({\mathcal H}_{j-i}^{n-1};W_{1+i}\big),
\end{eqnarray*}
where $W_i$ is the generator of the $i$th floor of the Lie triangle, $(\; ; \; )$ designates the Poisson bracket and $\left(\begin{array}{c}\!\!j\!\!\\\!\!i\!\!\end{array}\right)$ is the binomial coefficient. Note that the order (or the number of floors of the Lie triangle) is chosen in such a way that the transformation converges numerically, in other words we stop when we do not get more significant information by going one order further. \\
These generators will help us to compute the evolution of any variable. Let us show how to get the generator of the first floor. \\
We consider the first homological equation: ${\mathcal H}_0^1={\mathcal H}_1^0+({\mathcal H}_0^0;W_1)$. In this equation ${\mathcal H}_1^0$ is known and we choose ${\mathcal H}_0^1$ to be the average of ${\mathcal H}_1^0$. In other words, ${\mathcal H}_0^1$ will contain only terms without short periods and of order larger or equal to 3. So, expanding the Poisson bracket, and using equation (\ref{h00}), we have the following equation to solve:
\begin{eqnarray}
&&\underbrace{\frac{\partial {\mathcal H}_0^0}{\partial u}}_{0}\frac{\partial W_1}{\partial U}-\underbrace{\frac{\partial {\mathcal H}_0^0}{\partial U}}_{n_u}\frac{\partial W_1}{\partial u}+0-\underbrace{\frac{\partial {\mathcal H}_0^0}{\partial V}}_{n_v}\frac{\partial W_1}{\partial v}+0-\underbrace{\frac{\partial {\mathcal H}_0^0}{\partial W}}_{n_w}\frac{\partial W_1}{\partial w}+0-\underbrace{\frac{\partial {\mathcal H}_0^0}{\partial Z}}_{n_z}\frac{\partial W_1}{\partial z}={\mathcal H}_0^1-{\mathcal H}_1^0\nonumber \\
\Leftrightarrow&&-n_u\frac{\partial W_1}{\partial u}-n_v\frac{\partial W_1}{\partial v}-n_w\frac{\partial W_1}{\partial w}-n_z\frac{\partial W_1}{\partial z}={\mathcal H}_0^1-{\mathcal H}_1^0.
\end{eqnarray}
Since ${\mathcal H}_0^1$ only consists of all the terms of ${\mathcal H}_1^0$ without short periods, the right-hand side term of the previous equation only contains short periodic terms. It is then easy to compute $W_1$ and see that this is also only composed of short periodic terms. The computation of the other orders is done in a similar way.\\
With the generators, we can now compute the evolution of any function of the variables. First, we go back to cartesian coordinates to avoid the singularities when any of the moment is 0 (the angle is then undefined) and the formula is
\begin{equation}
f(x_1,x_2,\xi_1,\xi_2,y_1,y_2,\eta_1,\eta_2)=f(\bar x_1,\bar x_2,\bar \xi_1,\bar \xi_2,\bar y_1,\bar y_2,\bar \eta_1,\bar \eta_2)+\sum_{i=1}^{\text{order}}\frac{1}{i!}\left(f(x_1,x_2,\xi_1,\xi_2,y_1,y_2,\eta_1,\eta_2);W_i\right),
\end{equation}
where the Poisson bracket is evaluated at the equilibria $\bar x_1,\bar x_2,\bar \xi_1,\bar \xi_2,\bar y_1,\bar y_2,\bar \eta_1,\bar \eta_2$.\\
After the use of this Lie algorithm, we have our transformed Hamiltonian in the diagonal, without short periods: $\bar{\mathcal H}=\sum_{i=0}^{\text{order}} {\mathcal H}^i_0/i!$. \\
Until here, except for the fact that we have 2 additional degrees of freedom, our process is very similar to the one described in our previous study \citep{dnrl09}. The main difference is the change of fundamental frequencies.\\
In this Hamiltonian, the linear terms in $U$, $V$, $W$ and $Z$ changed with the transformation process, yielding corrections to the free frequencies. These corrections also appeared in our 2-degree of freedom work, but so imperceptibly that we did not mention it.\\
On the other hand, it plays a major role here. Here are the fundamental periods with the corrections:
\begin{eqnarray}
&& T_u=12.0568\text{ years,} \qquad T_v=1626.51\text{ years,}\\
&& T_w=337.853\text{ years,} \qquad T_z=58.6189\text{ days}.
\end{eqnarray}
We notice the very large change in $T_v$, the period related to the libration in latitude. \\
The period of rotation of Mercury is 58.646 days. The fundamental period related to the wobble of the core of Mercury is really close to this fundamental period and this particular combination of angles is present in the series related to the degree of freedom related to the libration in latitude. We are in fact close to a resonance that alters the efficiency of the algorithm.\\
As a consequence, the numerical convergence of the algorithm is really slow when we take values of $\epsilon_3$ close to $\epsilon_1$. For $\epsilon_3=\epsilon_1$, we must use a 13th order Lie triangle to barely converge to the free period $T_v$. During this process, we multiply series of millions of terms, which takes several days of computation. The process is divergent whenever $\epsilon_3\le 0.9\;\epsilon_1$. \\
Later in the paper, we draw a table of these periods for different values of $\epsilon_3$ and $\epsilon_4$ (cf. Table \ref{tab:propfreq}).

\subsection{Numerical study}

\par As we already did for the rotation of Mercury with a spherical core \citep{dnrl09} or for rigid bodies (see e.g. \citet{n10}), we performed a numerical study of the system by numerically integrating the equations derived from the Hamiltonian (\ref{equ:hamiltout}), and then carrying out a frequency analysis, for different values of the parameters $\epsilon_3$ and $\epsilon_4$.

\par In order to integrate numerically the system, we first express the coordinates of the Sun (x,y) (as in Eq.\ref{equ:pull2}) thanks to Poisson series given by the VSOP planetary theory and the rotations given in (Eq.\ref{equ:passage}). This way, we get coordinates depending on the canonical variables. Then we derive the equations coming from the Hamiltonian (\ref{equ:hamiltout}), such as:

\begin{eqnarray}
\frac{dp}{dt} =  \frac{\partial \mathcal{H}}{\partial P}, & &  \frac{dP}{dt} = -\frac{\partial \mathcal{H}}{\partial p}.
\end{eqnarray}
We then integrate over 13,000 years using the Adams-Bashforth-Moulton 10th order predictor-corrector integrator.

\par The initial conditions are chosen close to the equilibrium and are iteratively refined to get amplitudes of the free librations as small as possible (see e.g. \citet{dnrl09}). The reason is that we want to be able to extract the forced rotational motion of Mercury as accurately as possible, while the free terms are a source of noise. Moreover, these free terms are expected to be damped thanks to dissipations \citep{p05}. In addition to these forced terms, the proper frequencies are worth to be determined because of their significance on the dynamics of the system. We get their values from the first iteration.

\par  The frequency analysis algorithm that we use is based on Laskar's original idea, named NAFF as Numerical Analysis of the Fundamental Frequencies (see for instance \citet{l93} for the method, and \citet{l03} for the convergence proofs). It aims at identifying the coefficients $a_k$ and $\omega_k$ of a complex signal $f(t)$ obtained numerically over a finite time span $[-T;T]$  and verifying

\begin{equation}
\label{equ:naff}
f(t) \approx \sum_{k=1}^na_k\exp(\sqrt{-1}\omega_kt),
\end{equation}
where $\omega_k$ are real frequencies and $a_k$ complex coefficients. If the signal $f(t)$ is real, its frequency spectrum is symmetric and the complex amplitudes associated with the frequencies $\omega_k$ and $-\omega_k$ are complex conjugates. The frequencies and amplitudes associated are found with an iterative scheme. To determine the first frequency $\omega_1$, one searches for the maximum of the amplitude of 

\begin{equation}
\label{equ:philas}
\phi(\omega)=<f(t),\exp(\sqrt{-1}\omega t)>,
\end{equation}
where the scalar product $<f(t),g(t)>$ is defined by

\begin{equation}
\label{equ:prodscal}
<f(t),g(t)>=\frac{1}{2T}\int_{-T}^T f(t)g(t)^*\chi(t) dt,
\end{equation}
$g(t)^*$ being the complex conjugate of $g(t)$ .$\chi(t)$ is a weight function alike a Hann or a Hamming window, i.e. a positive function verifying

\begin{equation}
\label{equ:poids}
\frac{1}{2T}\int_{-T}^T \chi(t) dt=1.
\end{equation}
Using such a window can help the determination in reducing the amplitude of secondary minima in the transform (\ref{equ:prodscal}). Its use is optional.
\par Once the first periodic term $\exp(\sqrt{-1}\omega_1t)$ is found, its complex amplitude $a_1$ is obtained by orthogonal projection, and the process is started again on the remainder $f_1(t)=f(t)-a_1\exp(\sqrt{-1}\omega_1t)$. The algorithm stops when two detected frequencies are too close to each other, what alters their determinations, or when the number of detected terms reaches a limit set by the user. This algorithm is very efficient, except when two frequencies are too close to each other. In that case, the algorithm is not confident in its accuracy and stops. When the difference between two frequencies is larger than twice the frequency associated with the length of the total time interval, the determination of each fundamental frequency is not perturbed by the other ones. Although the iterative method suggested by \citet{c98} allows to reduce this distance, some troubles may remain. In our particular case, these problems are likely to arise because of the proximity between the free frequency of the core $\omega_z$ and the frequency of the spin.

\subsection{Results}

\par The Table \ref{tab:propfreq} gives our analytical and numerical results, for 5 different sets of values for $\epsilon_3$ and $\epsilon_4$, $\epsilon_1$, $\epsilon_2$ being fixed with the known values of $J_2$, $C_{22}$ and $\delta=1-C_m/C$. These 5 different sets are: 

\begin{itemize}

\item Case 1 : $\epsilon_3=\epsilon_4=0$. This is a singular case, because the core-mantle interactions should not exist when the core is spherical. Moreover, we are at the exact resonance between the proper mode $\omega_z$ and the spin frequency $\omega$. We computed this case to look for an agreement with our previous study \citep{dnrl09}, in which the spherical core was just removed.

\item Case 2 : $\epsilon_3/\epsilon_1=0.1$, $\epsilon_4=0$. We are close to the resonance, we here aim at detecting any discontinuity in the behaviour of the system close to the exact resonance.

\item Case 3 : $\epsilon_3/\epsilon_1=\epsilon_4/\epsilon_2=1$. The cavity is homothetical to Mercury, this was the configuration studied by \citet{h08} for Io.

\item Case 4 : $\epsilon_3/\epsilon_1=3$, $\epsilon_4=0$. We here take some distance with the resonance.

\item Case 5 : $\epsilon_3/\epsilon_1=\epsilon_4/\epsilon_2=3$. By comparing this configuration with the previous one, we study the influence of the parameter $\epsilon_4$, i.e. the equatorial ellipticity of the core.

\end{itemize}
We made tests for a wider range of the parameters, but retained only these 5 cases for sake of conciseness.

\begin{table}
\centering
\caption{Proper periods of the system, numerically and analytically determined with a good agreement. Analytical values are missing when $\epsilon_3/\epsilon_1$ is small, because of singularities met by our algorithm of Lie transforms. $\omega$ is the spin frequency of Mercury, the last line represents the distance of the system with the resonance between $\omega$ and $\omega_z$. We do not give any numerical value of $T_z$ for the exact resonance ($\epsilon_3=\epsilon_4=0$) because we actually cannot numerically distinguish the free librations of the core from forced contributions at $58.646$ days.\label{tab:propfreq}}
\begin{tabular}{c|r|r|rr|rr|rr}
\hline\hline
$\epsilon_3/\epsilon_1$ & $0$ & $0.1$ & \multicolumn{2}{c}{$1$} & \multicolumn{2}{c}{$3$} & \multicolumn{2}{c}{$3$}\\
$\epsilon_4/\epsilon_2$ & $0$ & $0$ & \multicolumn{2}{c}{$1$} & \multicolumn{2}{c}{$3$} & \multicolumn{2}{c}{$0$} \\
 & Numerical & Numerical & Analytical & Numerical & Analytical & Numerical & Analytical & Numerical \\
\hline
$T_u$ (y) & $12.05800$ & $12.05775$ & $12.05685$ & $12.05772$ & $12.05685$ & $12.05777$ & $12.05685$ & $12.05773$ \\
$T_v$ (y) & $615.77$   & (large)      & $1626.51$ & $1636.43$  & $1216.46$ & $1214.91$  & $1216.41$ & $1216.09$ \\
$T_w$ (y) & $337.82$   & $337.82$   & $337.85$ & $337.87$   & $338.03$ & $338.14$   & $338.03$ & $338.20$ \\
$T_z$ (d) &  --  & $58.630$   & $58.619$ & $58.619$   & $58.585$ & $58.585$ & $58.585$ & $58.585$ \\
$T_{z-\omega}$ (y) & -- & $574.06$ & $344.88$ & $343.45$ & $154.08$ & $154.04$ & $154.05$ & $154.01$ \\
\hline
\end{tabular}
\end{table}

\par This table gives us first results on the influence of the shape of the core on the behaviour of the system. We can notice that the two periods $T_u$ and $T_w$ are quite constant. This is all the most interesting for $T_u$ because this degree of freedom, i.e. the longitudinal motion, is very weakly coupled with the variations of the obliquity, and even less with the two other ones. This means that this longitudinal motion is basically not influenced by the shape of the core, so the amplitude of longitudinal librations should depend only on $C_m/C$, even with very accurate observations.

\par
\begin{figure}
\centering
\includegraphics{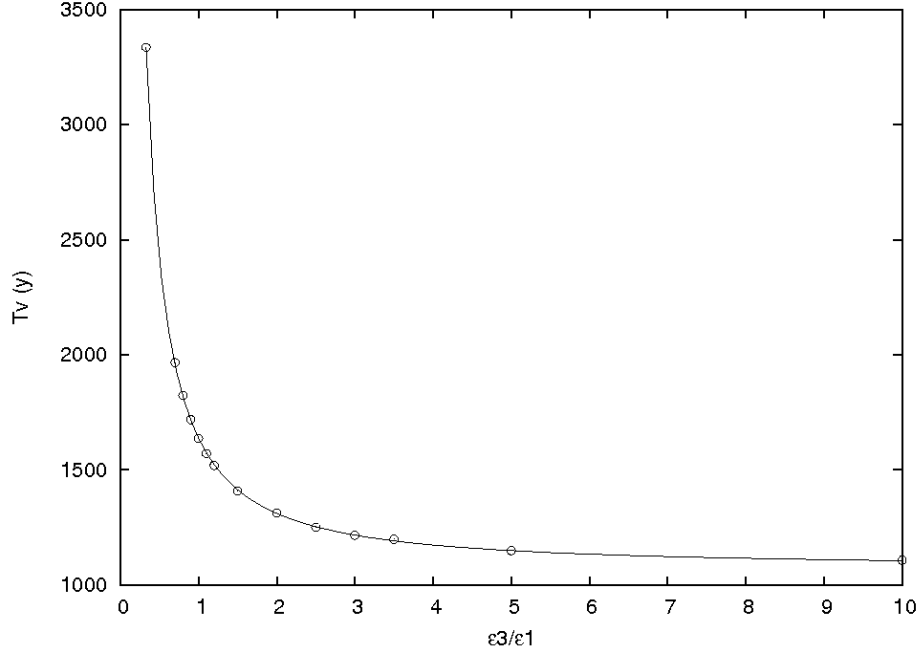}
\caption{Graphical representation of the proper period $T_v$, associated with the obliquity. This curve can be fitted by $f(\epsilon_3)\approx T_{vr}+A\epsilon_3^{-5/4}$, where $A$ is a constant and $T_{vr}$ is the value of $T_v$ when Mercury is considered as rigid. The circles are the outputs of the frequency analysis after numerical integration, while the solid line is $f(\epsilon_3/\epsilon_1)=A(\epsilon_3/\epsilon_1)^B+C$ with $A=564.488$ y, $B=-1.25224$ and $C=1,074.3$ y.\label{fig:graphtv}}
\end{figure}

 The variations of the periods $T_v$ are worth noticing, because they present a discontinuous behaviour. From $\epsilon_3/\epsilon_1=3$ to $0.1$ this period is getting larger and larger, some tests at $\epsilon_3/\epsilon_1=0.33$ giving $T_v=3,335.16$ years, while this period is too long to be numerically determined for $\epsilon_3/\epsilon_1=0.1$. However, we have $T_v \approx 616$ years at the exact resonance, what is consistent with the results obtained by simply removing the spherical core from the system. We think that it emphasizes the change of behaviour when the system is trapped into the resonance. As it can be seen in Table \ref{tab:periodnum} and Figure \ref{fig:graphtv}, out of the resonance the period is decreasing, tending to reach the rigid value of $1,065$ years. A least-square fit of $T_v$ gives $T_v\approx A(\epsilon_3/\epsilon_1)^B+C$, with $A=564.488\pm4.146$ y, $B=-1.25224\pm6.003\times10^{-3}$ and $C=1,074.3\pm3.233$ y. By including $\epsilon_1$ in the constant and considering that $1,074$ is close to the rigid value of $1,065$ years \citep{dl04,rb04}, we can guess for $T_v$ an evolution such as

\begin{equation}
T_v \approx A\epsilon_3^{-5/4}+T_{vr},
\label{equ:Tvapp}
\end{equation}
where $A$ is a constant and $T_{vr}$ is the rigid value of $T_v$. The influence of $\epsilon_4$ is very small.

\begin{table}
\caption{Evolution of the periods $T_v$ and $T_z$ with respect to $\epsilon_3/\epsilon_1$, with $\epsilon_4=0$. These periods have been numerically determined, and confirmed analytically with a very good agreement when $\epsilon_3/\epsilon_1>1$. \label{tab:periodnum}}
\centering
\begin{tabular}{l|rrr}
\hline\hline
$\epsilon_3/\epsilon_1$ & $T_v$ (y) & $T_z$ (d) & $T_{z-\omega}$ (y) \\
\hline
$0.33$ & $3335.16$ & $58.628$ & $511.17$ \\
$0.7$  & $1966.31$ & $58.623$ & $409.08$ \\
$0.8$  & $1823.63$ & $58.622$ & $385.35$ \\
$0.9$  & $1718.34$ & $58.620$ & $363.50$ \\
$1.0$  & $1636.35$ & $58.619$ & $343.46$ \\
$1.1$  & $1570.86$ & $58.617$ & $325.10$ \\
$1.2$  & $1519.36$ & $58.616$ & $308.30$ \\
$1.5$  & $1408.10$ & $58.611$ & $266.01$ \\
$2$    & $1313.11$ & $58.602$ & $214.85$ \\
$2.5$  & $1250.26$ & $58.594$ & $179.64$ \\
$3$    & $1216.09$ & $58.585$ & $154.01$ \\
$3.5$  & $1198.68$ & $58.576$ & $134.72$ \\
$5$    & $1149.35$ & $58.550$ & $97.69$ \\
$10$   & $1107.62$ & $58.462$ & $50.83$ \\
\hline
\end{tabular}
\end{table}

\par In the same way as $T_v$, $T_z$ is decreasing when $\epsilon_3$ is growing. It confirms that the resonance between the proper rotation of the core and the period of the spin is reached for $\epsilon_3=0$. The third column of Table \ref{tab:periodnum} is more significant because it represents the distance of this frequency $\omega_z$ from the exact resonance.

\section{Consequences on the observable rotation}

\par Our canonical variables are very convenient to describe the dynamics of the system, unfortunately they are not observable variables. Observations of the rotation of Mercury are in fact observations of its surface, i.e. the rigid mantle in our model. So, we have to express the components of the angular momentum of the mantle $\vec{N^m}=A^m\omega_1\vec{f_1}+B^m\omega_2\vec{f_2}+C^m\omega_3\vec{f_3}$.

\subsection{The observable variables}

\par We deduce from the equations (\ref{equ:N1}) and (\ref{equ:N1c}):

\begin{equation}
\omega_1=\frac{D_1N^c_1+A^cN_1}{D_1^2-AA_c}
\label{equ:omega1}
\end{equation}
and 

\begin{equation}
\nu_1=\frac{D_1N_1+AN^c_1}{D_1^2-AA_c},
\label{equ:nu1}
\end{equation}
and similar formulae for $\omega_2$, $\omega_3$, $\nu_2$ and $\nu_3$.

\par We can now easily deduce the expression of the angular momentum of the mantle with respect to the components of $\vec{N}$ and $\vec{N^c}$:

\begin{equation}
\vec{N^m}  =  A^m\frac{D_1N^c_1-A^cN_1}{D_1^2-AA_c}\vec{f_1}+B^m\frac{D_2N^c_2-B^cN_2}{D_2^2-BB_c}\vec{f_2}+C^m\frac{D_3N^c_3-C^cN_3}{D_3^2-CC_c}\vec{f_3}.
\label{equ:Nm}
\end{equation}

\par We can define a wobble $J_m$ and a precession angle $l_m$ of the mantle of Mercury this way:

\begin{equation}
\vec{N^m}  =  G^m\sin J_m\sin l_m\vec{f_1}+G^m\sin J_m \cos l_m\vec{f_2}+G^m\cos J_m\vec{f_3},
\label{equ:Nmwob}
\end{equation}
where $G^m$ is the norm of $\vec{N^m}$. Because of the $3:2$ spin-orbit resonance, $G^m$ is expected to be close to $3nC^m/2$. We get $G^m$, $J_m$ and $l_m$ in equating the equations (\ref{equ:Nm}) and (\ref{equ:Nmwob}).

\par We now need to express the obliquity $K_m$ and the node $h_m$ of the mantle relatively to the inertial frame $(\vec{e_1},\vec{e_2},\vec{e_3})$. Naming $T_1$, $T_2$ and $T_3$ the coordinates of $\vec{N^m}$ in the inertial frame, we have:

\begin{equation}
\left(\begin{array}{c}
T_1 \\
T_2 \\
T_3
\end{array}\right)
=R_3(h_m)R_1(K_m)R_3(g_m)\left(\begin{array}{c}
0 \\
0 \\
G^m
\end{array}\right)
\label{equ:passag}
\end{equation}
and we get:

\begin{eqnarray}
T_1 &=& G^m\sin(K_m)\sin(h_m), \label{equ:t1} \\
T_2 &=& -G^m\sin(K_m)\cos(h_m), \label{equ:t2} \\
T_3 &=& G^m\cos(K_m). \label{equ:t3}
\end{eqnarray}

\par Naming $T_1'^c$, $T_2'^c$ and $T_3'^c$ the coordinates of the angular momentum of the core $\vec{N'^c}$ in the inertial frame $(\vec{e_1},\vec{e_2},\vec{e_3})$, we have:

\begin{equation}
\left(\begin{array}{c}
T_1'^c \\
T_2'^c \\
T_3'^c
\end{array}\right)
=R_3(h)R_1(K)R_3(g)R_1(J)R_3(l)\left(\begin{array}{c}
N_1'^c \\
N_2'^c \\
N_3'^c
\end{array}\right),
\label{equ:passage2}
\end{equation}
we now get:

\begin{eqnarray}
G^m\sin(K_m)\sin(h_m)&=&G\sin(K)\sin(h)-T_1'^c, \label{equ:km1} \\
G^m\sin(K_m)\cos(h_m)&=&G\sin(K)\cos(h)+T_2'^c, \label{equ:km2} \\
G^m\cos(K_m)&=&G\cos(K)-T_3'^c, \label{equ:km3}
\end{eqnarray}
and finally

\begin{eqnarray}
K_m & = & \arccos\Big(\frac{G\cos(K)-T_3'^c}{G^m}\Big), \label{equ:Km} \\
h_m & = & \arctan\Big(\frac{G\sin(K)\sin(h)-T_1'^c}{G\sin(K)\cos(h)+T_2'^c}\Big). \label{equ:hm}
\end{eqnarray}
From (Eq.\ref{equ:passag}), (Eq.\ref{equ:Km}) and (Eq.\ref{equ:hm}), we can deduce $g_m$. It is now straightforward to derive an angle $p_m$, analogous to $p$, with

\begin{equation}
\label{equ:pm}
p_m=l_m+g_m+h_m.
\end{equation}
The wobble $J'_c$ and the precession angle $l'_c$ of the core are directly derived from (Eq.\ref{equ:NPc2}), in the same way as $J_m$ and $l_m$. They are not observable variables, but can be of planetological interest.

\subsection{Results}

\par The outputs given are chosen for their physical relevance. We express the longitudinal motion of Mercury, the obliquity of the mantle with respect to its orbital motion, the polar motion of the mantle and the wobble of the core. The first three variables can be observed, while the last one has indirect implications like on Mercury's magnetic field.

\subsubsection{Longitudinal motion}

\par The longitudinal librations have been studied for a spherical core \citep{dnrl09,pmy09}. These studies show two main contributions: a $88$-day one of around $36$ arcsec, due to the variations of the Sun-Mercury distance because of Mercury's eccentricity, and a $11.86$-year one of approximately $42$ arcsec, due to the Jovian perturbation on Mercury's orbit. Its amplitude is highly sensitive to the size of the core, because of the proximity of a secondary resonance with the proper frequency $\omega_u$, the period associated being $12.06$ years. We should keep in mind that the Jovian contribution has actually not been observed. Considering the uncertainty on the ratio $(B-A)/C_m=(2.033\pm0.114)\times10^{-4}$, we can only say that the amplitude associated in the longitudinal motion of Mercury's mantle is bigger than $15$ arcsec (see e.g. \citet{ymp10}, Fig.5).

\par We have determined the longitudinal librations thanks to a frequency analysis of the angle $p_m$ (Eq.\ref{equ:pm}), after removal of a slope, i.e. the spin frequency of $39.1318408$ rad/y. In all our numerical simulations, we get an amplitude between $35.82915$ and $35.830$ arcsec for the $88$-day contribution, and between $41.316$ and $41.321$ arcsec for the $11.86$-year one. These results are in good agreement with the previous studies considering a spherical core, and do not show any significant variations. \citet{rvdb07} had the same conclusion in a similar study, this means that the shape of the liquid (or molten) core cannot be derived from observations of the longitudinal motion. This result could be expected from the negligible variations of the proper period associated, i.e. $T_u$ (Table \ref{tab:propfreq}).

\subsubsection{The obliquity of the mantle}

\par
\begin{figure}
\centering
\begin{tabular}{cc}
\includegraphics[height=5cm,width=7.8cm]{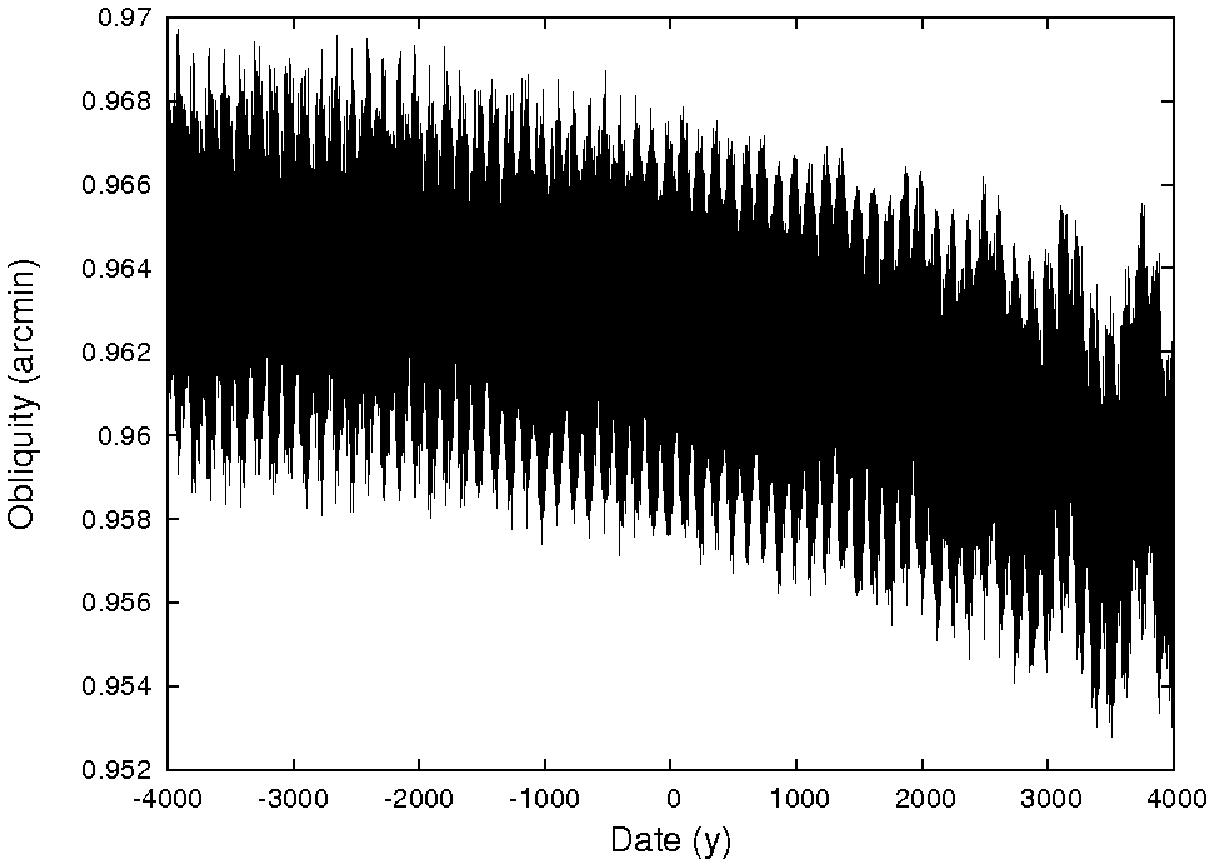} & \includegraphics[height=5cm,width=7.8cm]{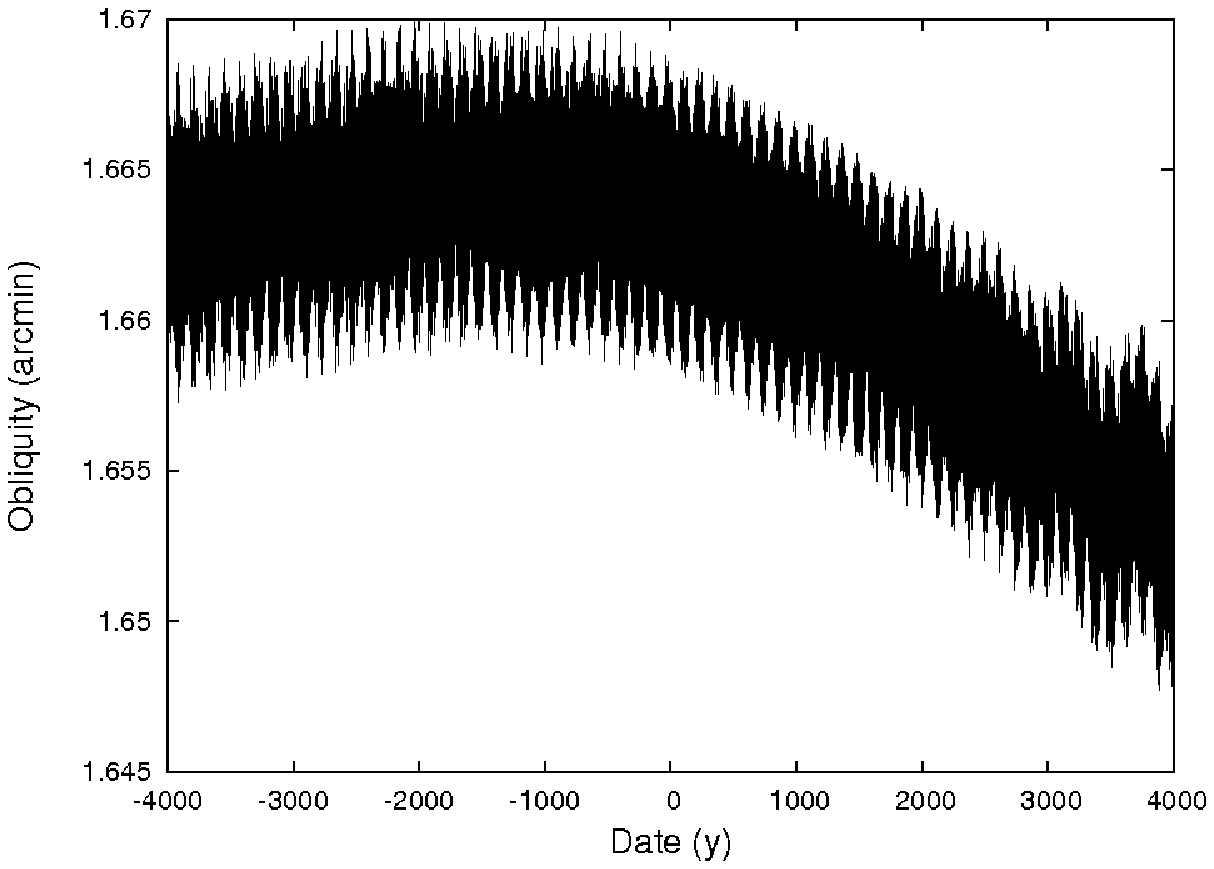} \\
$\epsilon_3=\epsilon_4=0$ & $\epsilon_3=\epsilon_1/10,\epsilon_4=0$ \\
\includegraphics[height=5cm,width=7.8cm]{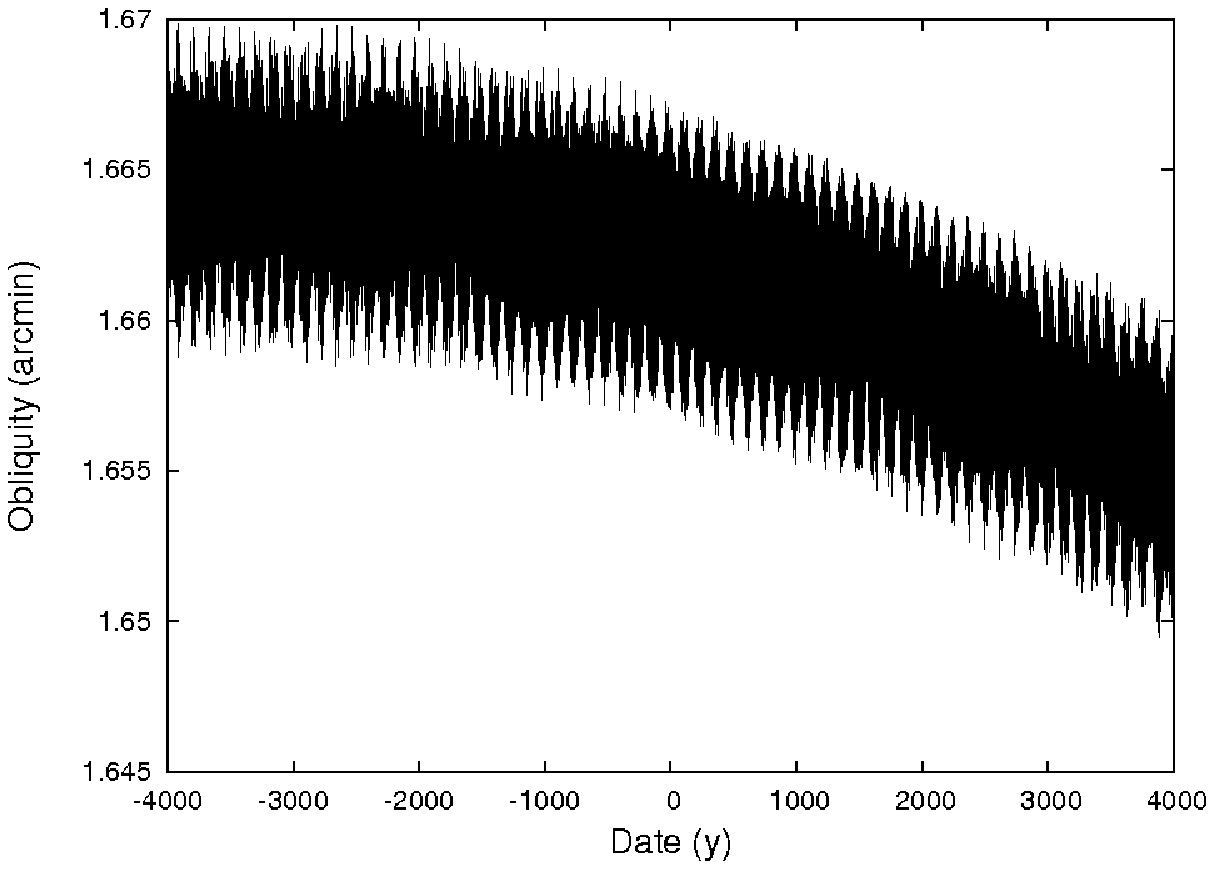} & \includegraphics[height=5cm,width=7.8cm]{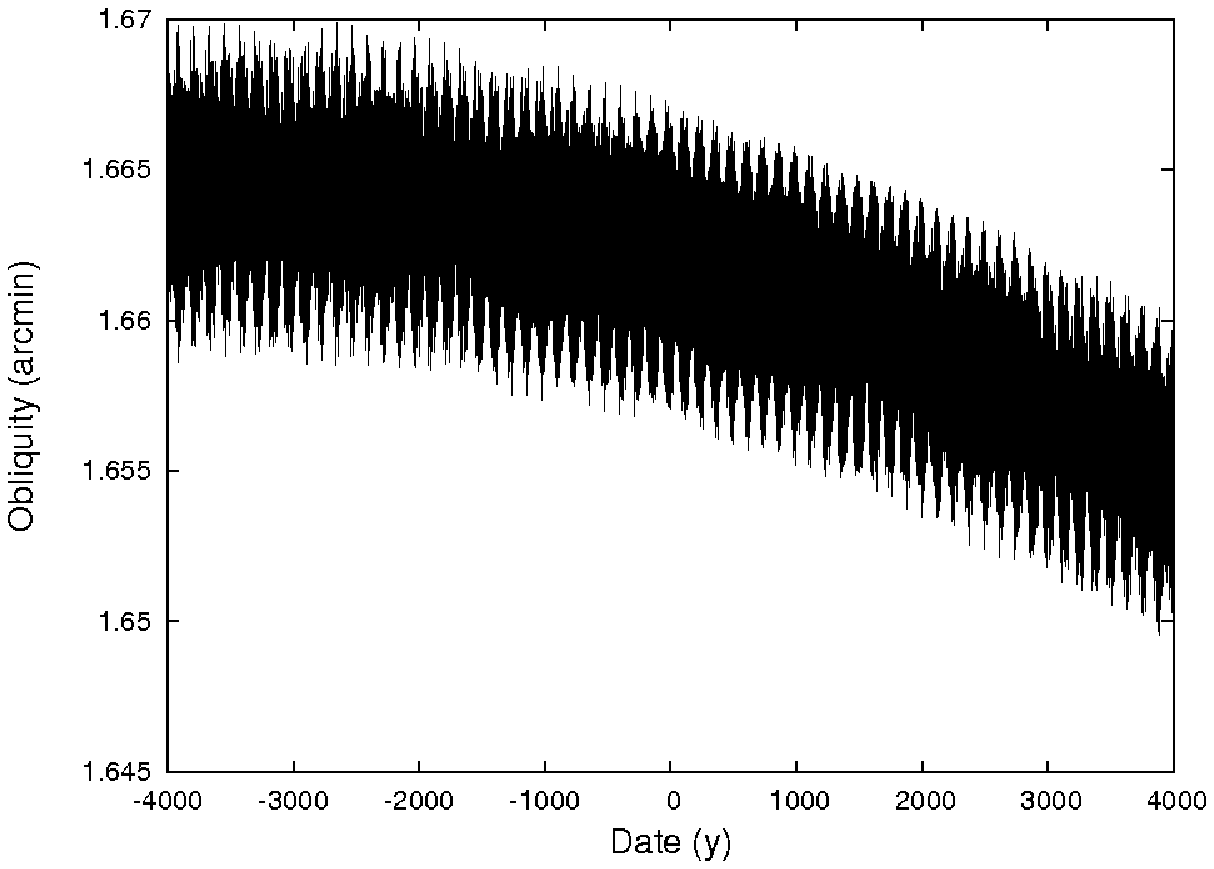} \\
$\epsilon_3/\epsilon_1=\epsilon_4/\epsilon_2=1$ & $\epsilon_3=3\epsilon_1,\epsilon_4=0$ \\
\end{tabular}
\caption{Variations of the orbital obliquity $\epsilon$ of the mantle of Mercury for different shapes of the core. The time origin is J2000. \label{fig:obli}}
\end{figure}

 There are many ways to define the obliquity of the mantle. $K_m$ (Eq.\ref{equ:Km}) is the obliquity with respect of the inertial reference plane, i.e. the ecliptic at J2000. This quantity lacks of physical relevance, that is the reason why we here prefer to express the orbital obliquity $\epsilon$, i.e. the obliquity with respect to the normal to the orbit. It is derived from the scalar product between $\vec{N^m}$ and the normal to the orbit, given by the cross product of the position and velocity vectors of Mercury. We show it in Figure \ref{fig:obli}.

\par This quantity shows essentially a secular behaviour, as expected (see e.g. \citet{p06,ym06,dl08,dnrl09}). Moreover, we can see from the plots that the obliquity is always close to $1.66$ arcmin, except in the strict resonant case ($\epsilon_3=\epsilon_4=0$), where the obliquity is close to $1$ arcmin. So, we can consider that there are two possible different behaviours: the resonant, and the non-resonant one. The resonant case is very improbable because it requires a strict fine tuning of the parameter $\epsilon_3$ (related to the polar flattening of the core). So, we can say that the shape of the core could probably not be derived from observations. As stated for instance in \citep{p06}, these values of $1$ and $1.66$ arcmin depend on the value of $J_2$, on which we have a $30\%$ uncertainty.

\par We here consider the assumption of the liquid core to be valid at any timescale. It is in fact often stated that Mercury should be considered as rigid, for the adiabatic evolution of the obliquity \citep{ppssz02}. Under this last assumption, the expected obliquity should be about twice larger, as the $2.11\pm0.1$ arcmin measured by \citet{mpjsh07}.

\subsubsection{Polar motion of the mantle}

\par We give here the polar motion of the rotation axis of the mantle about the geometrical North Pole. Following \citet{h05}, we define the first two components $Q_1$ and $Q_2$ of the unit vector in the direction of the instantaneous axis of rotation by $Q_1 \approx \sin J_m \sin l_m [1+(C_m-A_m)/C_m]$ and $Q_2 \approx \sin J_m \cos l_m [1+(C_m-B_m)/C_m]$, and multiply them by the polar radius of Mercury, i.e. $2,439.7$ km \citep{saacch07}.

\par
\begin{figure}
\centering
\begin{tabular}{cc}
\includegraphics[height=5cm,width=7.8cm]{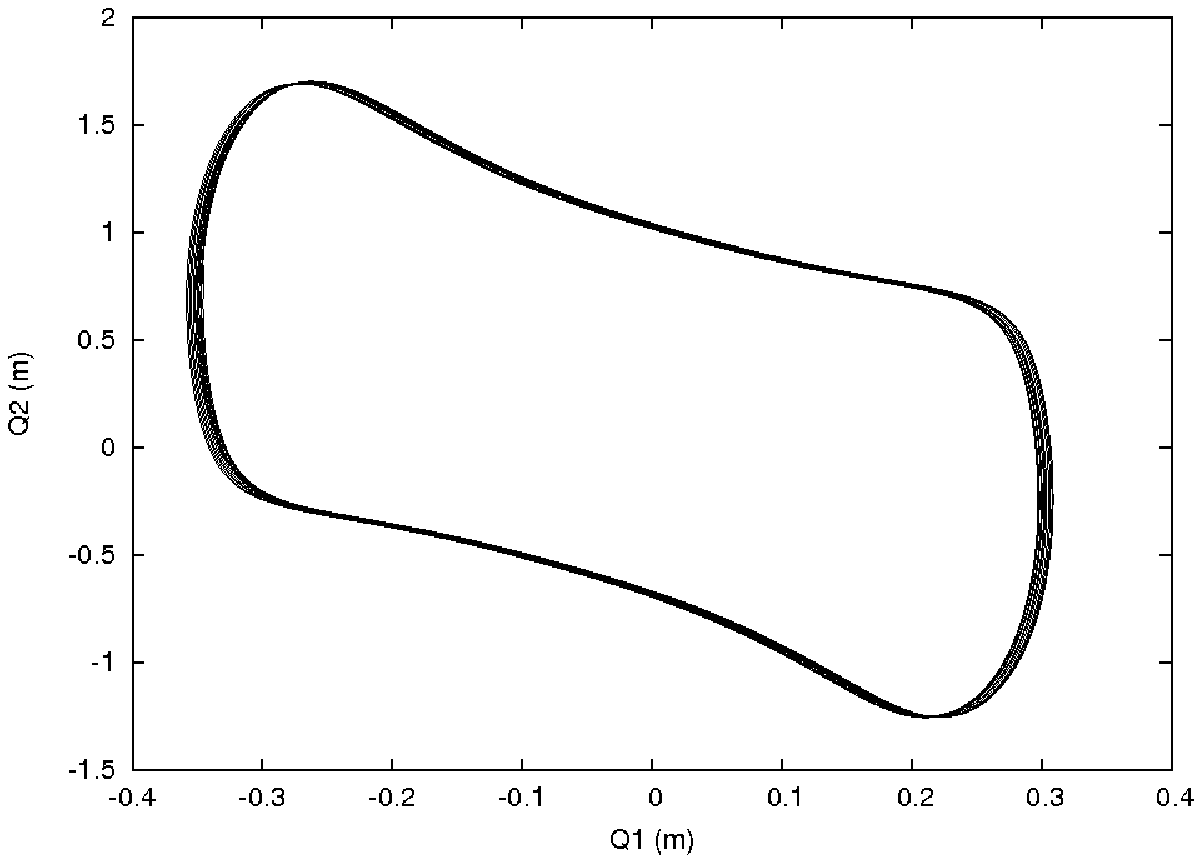} & \includegraphics[height=5cm,width=7.8cm]{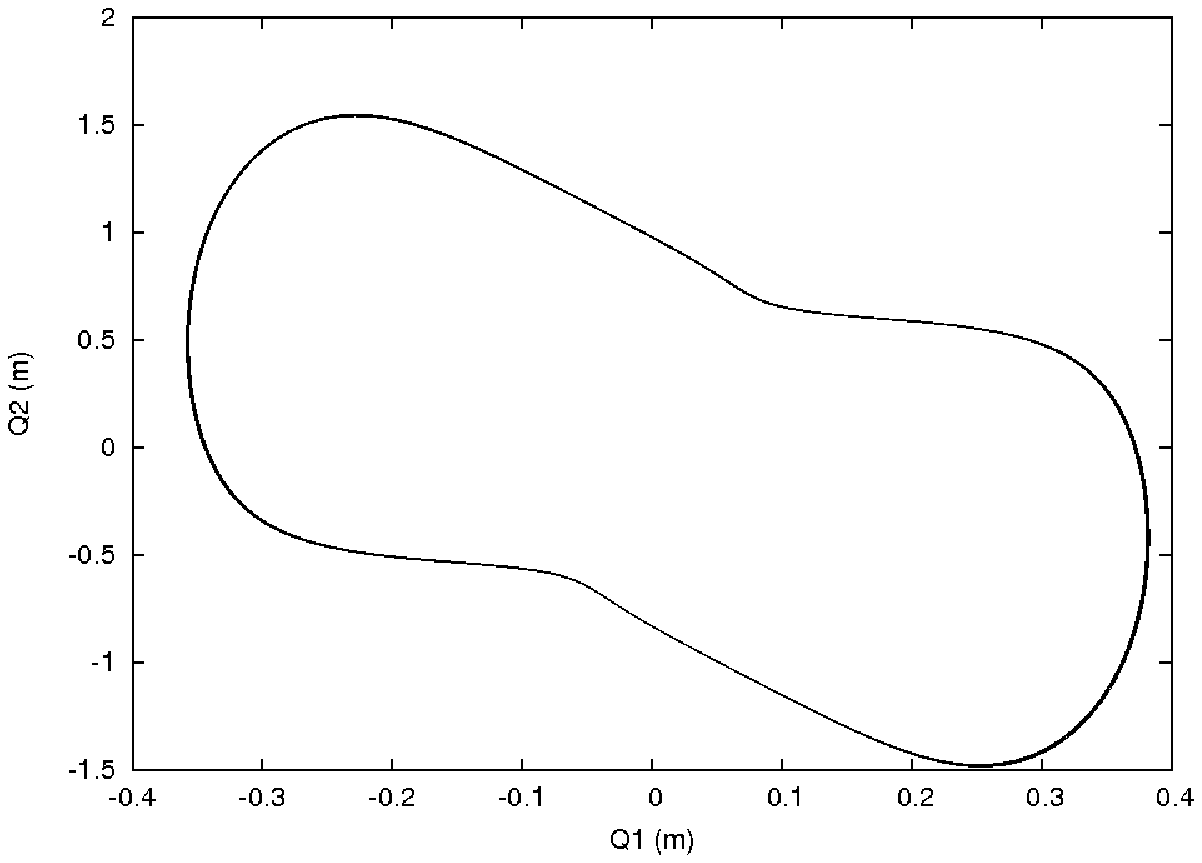} \\
$\epsilon_3/\epsilon_1=\epsilon_4/\epsilon_2=1$ & $\epsilon_3=3\epsilon_1,\epsilon_4=0$ \\
\end{tabular}
\caption{Polar motion of the mantle, plotted over 5 years from J2000. \label{fig:polarmantle}}
\end{figure}

 In Figure \ref{fig:polarmantle}, we show the polar motion of the mantle over 5 years, starting from J2000, for two different shapes of the core. In both cases, we see that this motion has a very small amplitude (smaller than 2 meters), and so should probably not be detected by the spacecrafts. They present similar aspects, the differences between the two representations being emphasized in Table \ref{tab:polarmantle}.

\par This table gives quasiperiodic representations of the polar motion of the mantle, defined by the quantity $Q_1+\sqrt{-1}Q_2$, in the same cases as in Figure \ref{fig:polarmantle}. The frequencies and the amplitudes associated have been numerically obtained using frequency analysis. The basic frequency of this motion, here named $\tau$, is the frequency of the sidereal hermean day. Its period is twice the orbital period of Mercury and thrice its rotational one. We can consider it as the frequency of precession of the rotation axis of the mantle about the geometrical pole. The other frequencies are harmonics of $\tau$. This table confirms that the amplitude of this motion is small. We can in particular notice that we cannot detect any resonant excitation of the $3\tau=\omega$ contribution, while it is close to the free frequency $\omega_z$.

\begin{table}
\centering
\caption{Synthetic representation of the polar motion of the mantle $Q_1+\sqrt{-1}Q_2$. This motion can be expressed with a high accuracy (see the amplitudes) just with harmonics of the sidereal hermean frequency $\tau$. The two columns "Amplitude" are related to the two cases shown in Figure \ref{fig:polarmantle}. We can see that the differences between the two cases are quite small. \label{tab:polarmantle}}
\begin{tabular}{rrrr}
\hline\hline
$\tau$ & Amplitude (m) & Amplitude (m) & Period \\
 & $\epsilon_3/\epsilon_1=\epsilon_4/\epsilon_2=1$ & $\epsilon_3=3\epsilon_1,\epsilon_4=0$ & (d) \\
\hline
 $1$ & $0.73344$ & $0.73118$ &  $175.9$ \\
$-1$ & $0.44199$ & $0.44064$ & $-175.9$ \\
 $3$ & $0.20077$ & $0.24330$ &   $58.6$ \\
$-3$ & $0.17937$ & $0.17844$ &  $-58.6$ \\
 $5$ & $0.06283$ & $0.06264$ &   $35.2$ \\
$-5$ & $0.06246$ & $0.06231$ &  $-35.2$ \\
 $7$ & $0.01765$ & $0.01760$ &   $25.1$ \\
$-7$ & $0.01490$ & $0.01486$ &  $-25.1$ \\
 $9$ & $0.00507$ & $0.00505$ &   $19.5$ \\
$-9$ & $0.00146$ & $0.00377$ &  $-19.5$ \\
\hline
\end{tabular}
\end{table}

\subsubsection{Polar motion of the core}

\par
\begin{figure}
\centering
\begin{tabular}{cc}
\includegraphics[height=5cm,width=7.8cm]{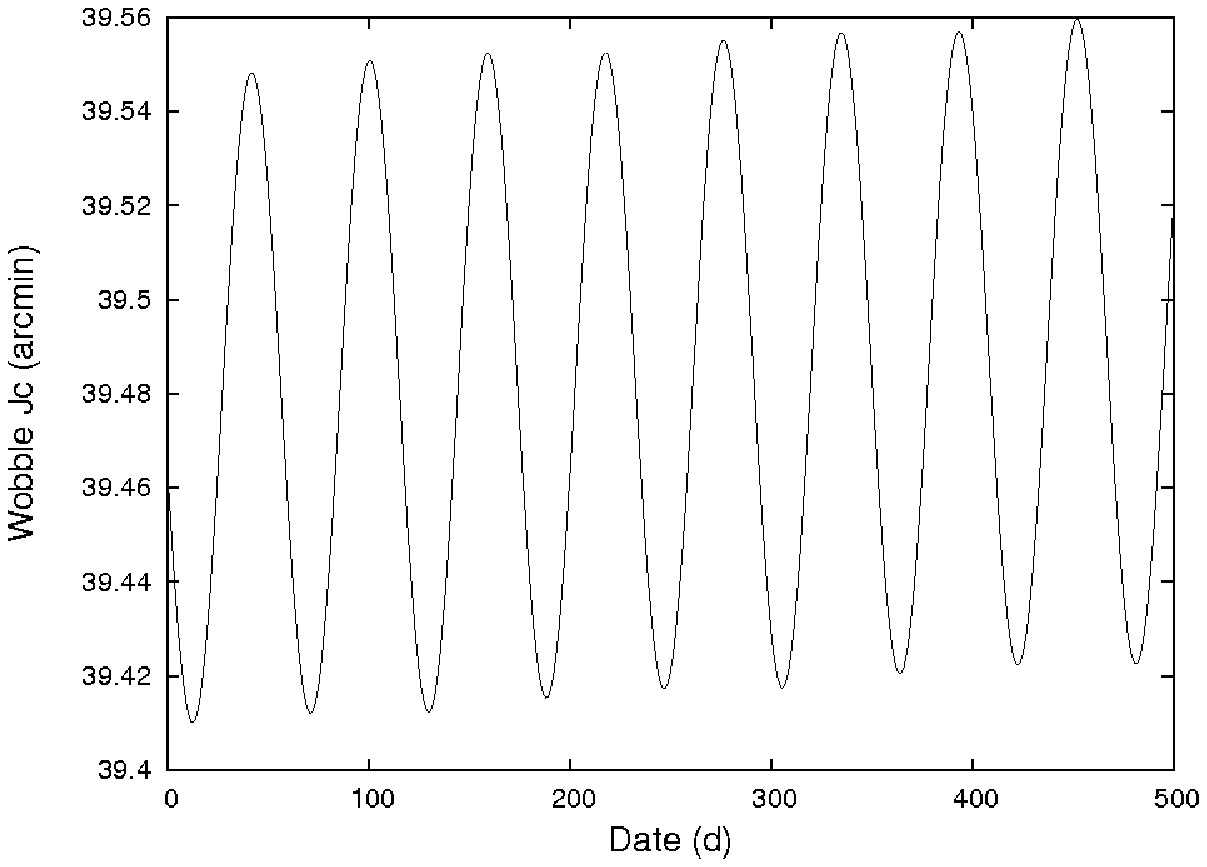} & \includegraphics[height=5cm,width=7.8cm]{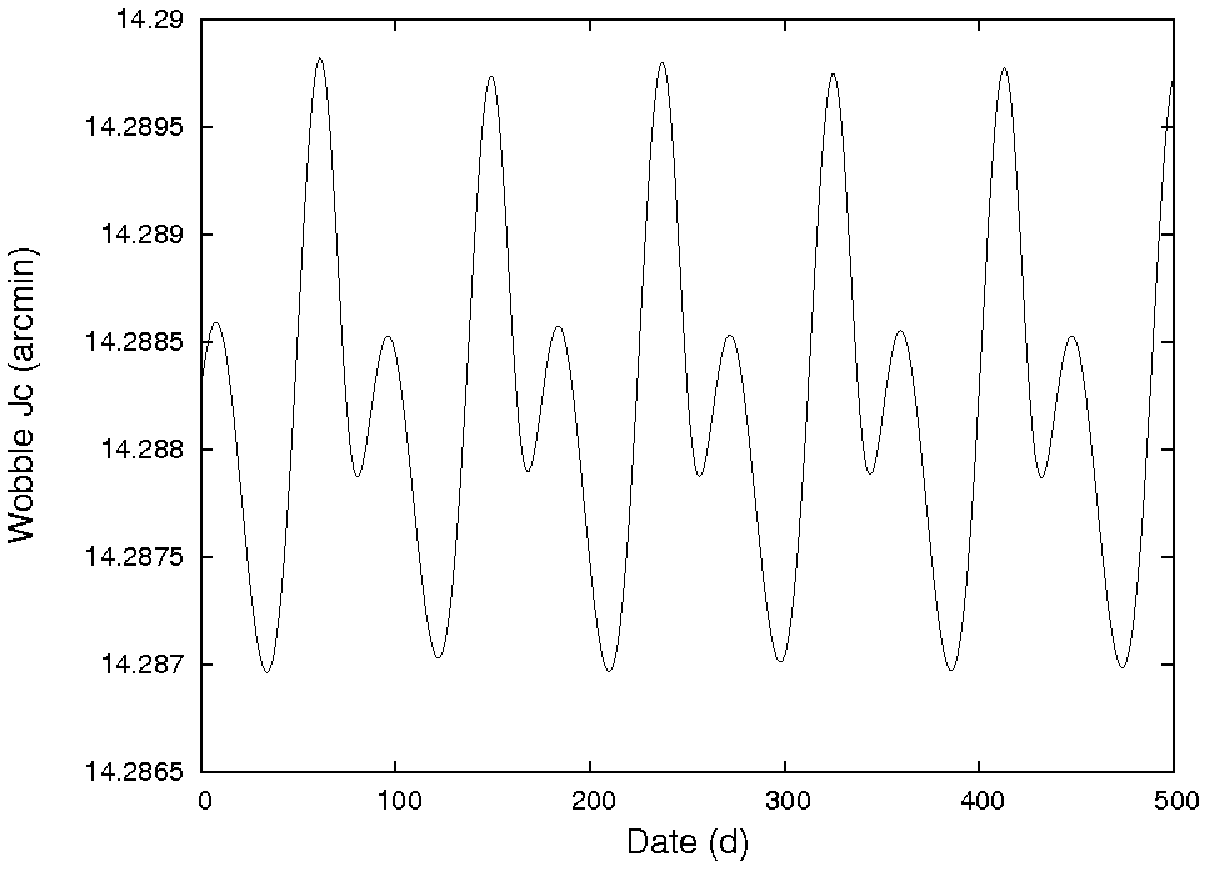} \\
$\epsilon_3=\epsilon_4=0$ & $\epsilon_3=\epsilon_1/10,\epsilon_4=0$ \\
\includegraphics[height=5cm,width=7.8cm]{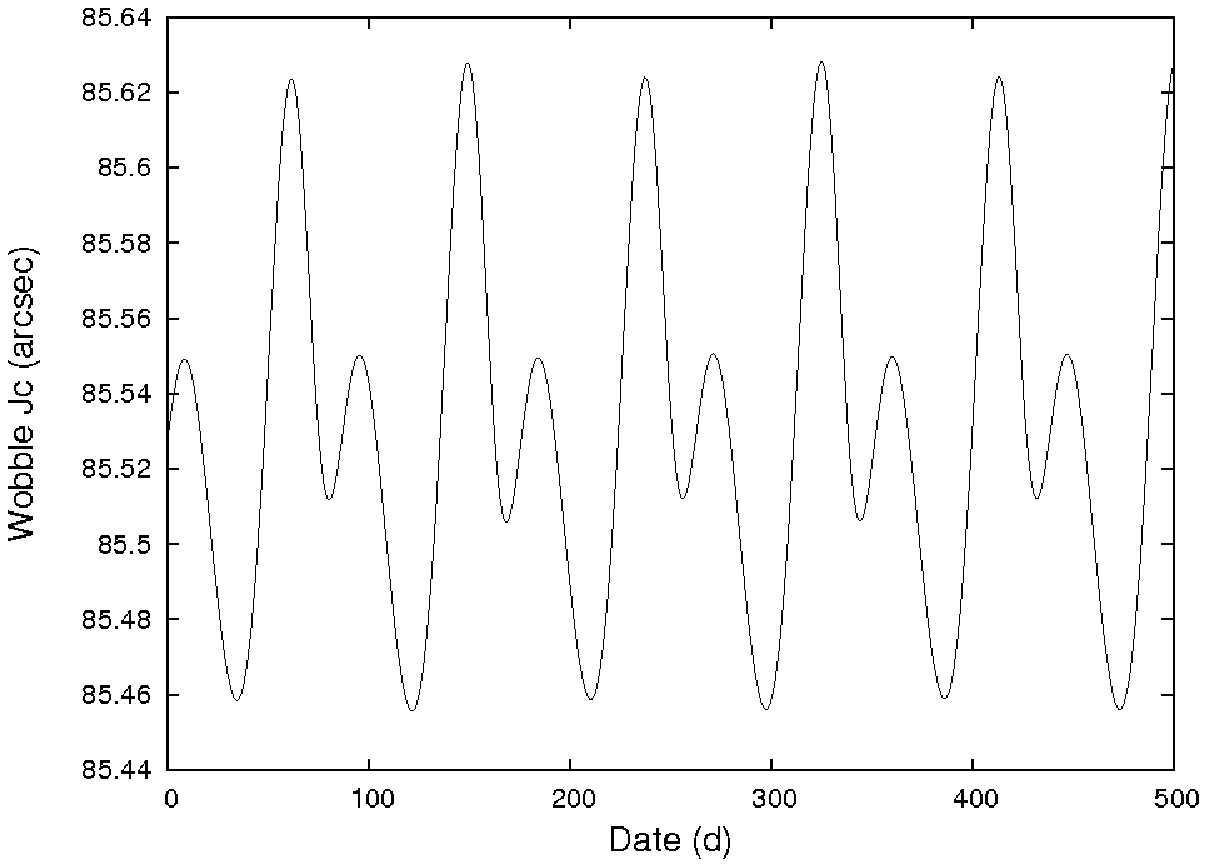} & \includegraphics[height=5cm,width=7.8cm]{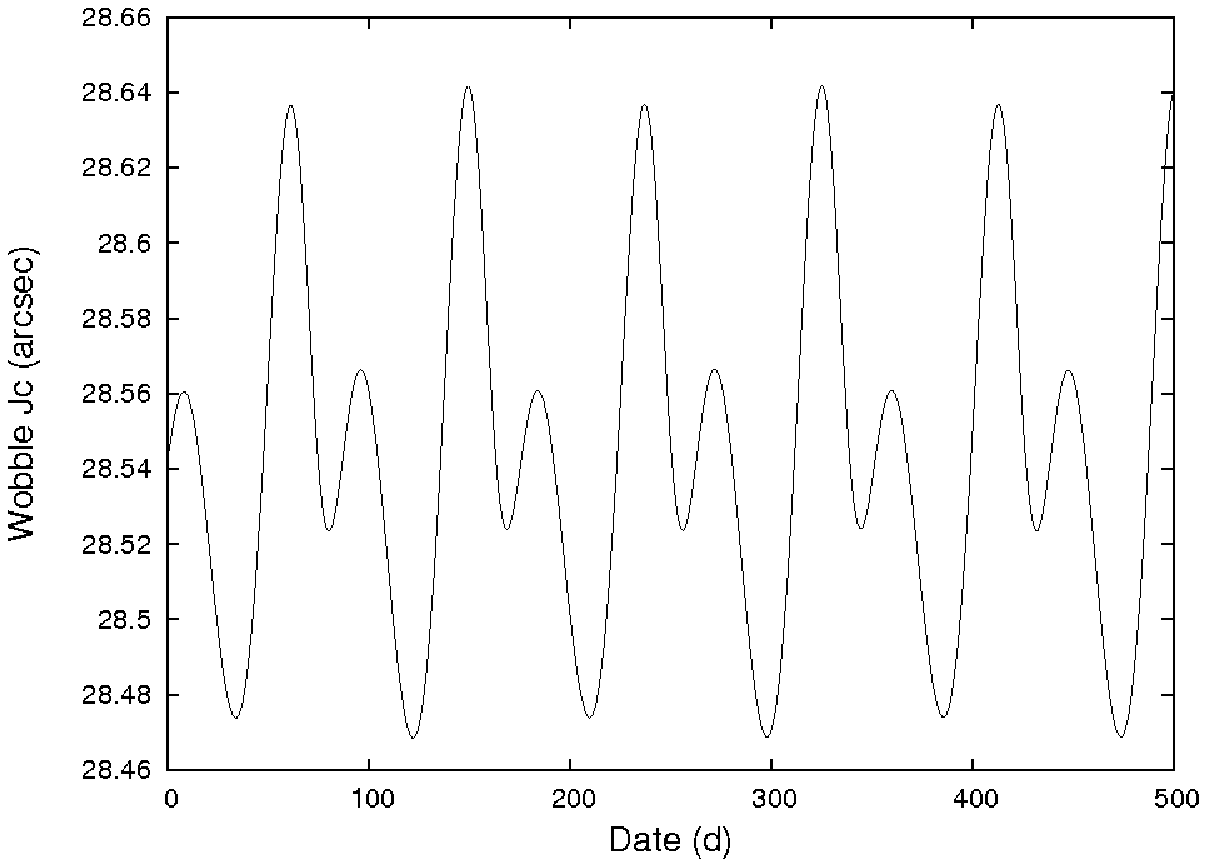} \\
$\epsilon_3/\epsilon_1=\epsilon_4/\epsilon_2=1$ & $\epsilon_3=3\epsilon_1,\epsilon_4=0$ \\
\end{tabular}
\caption{Wobble of the core $J_c$, obtained from numerical simulations after removal of the free librations. We can see that for a spherical core ($\epsilon_3=\epsilon_4=0$) the visual aspect is very different from the other cases. Moreover, a long-term visualisation of $J_c$ shows a slope, i.e. a secular increase of the wobble of the core, while the other cases (out of the resonance) do not show it.\label{fig:polarcore}}
\end{figure}

 Even if the core cannot be observed, its rotation should still be described. It could indeed have some planetological implications, as for instance the origin of Mercury's magnetic field (see e.g. \citet{c06}). Figure \ref{fig:polarcore} gives the evolution of its wobble $J_c$ for different values of the shape parameters $\epsilon_3$ and $\epsilon_4$, while Table \ref{tab:polarcore} is a synthetic description of the quantity $J_c\exp(\sqrt{-1}l_c)$, representing the precessional motion of the core.

\par The wobble $J_c$ is quite large when the system is close to the resonance, and its amplitude decreases when $\epsilon_3$ (the polar flattening of the core) increases, i.e. when the system takes distance from the exact resonance. In fact, for $\epsilon_3=0$, the behaviour of $J_c$ is a growing slope, meaning that the exact resonance tends to increase it dramatically, so the value of $40$ arcmin that can be read on Figure \ref{fig:polarcore} is not reliable. However, we are confident in the behaviour of the system out of the resonance, and we can see on the other three plots a quite similar visual aspect, except of the mean value. This aspect can be better understood thanks to Table \ref{tab:polarcore}.

\begin{table}
\centering
\caption{Synthetic representation of the precession of the rotation axis of the core about its geometrical pole axis. The quantity here analysed is $J_c\exp(\sqrt{-1}l_c)$. Contrary to the polar motion of the mantle, we see a high 58.6-d-contribution. The amplitude associated is raised by the proximity of the resonance with the proper frequency of the core.\label{tab:polarcore}}
\begin{tabular}{rrrrr}
\hline\hline
$\tau$ & Amplitude (arcsec) & Amplitude (arcsec) & Amplitude (arcsec) & Period \\
 & $\epsilon_3=\epsilon_1/10,\epsilon_4=0$ & $\epsilon_3/\epsilon_1=\epsilon_4/\epsilon_2=1$ & $\epsilon_3=3\epsilon_1,\epsilon_4=0$ & (d) \\
\hline
 $3$ & $858.344$ & $85.532$ & $28.546$ &   $58.6$ \\
$-3$ &         - &  $0.046$ &  $0.010$ &  $-58.6$ \\
 $1$ &        -  &  $0.049$ &  $0.053$ &  $175.9$ \\
$-1$ &   $0.045$ &  $0.011$ &  $0.046$ & $-175.9$ \\
 $5$ &   $0.067$ & - & $0.005$ &   $35.2$ \\
$-5$ &   - & - & $0.002$ &   $35.2$ \\
\hline
\end{tabular}
\end{table}

\par We see on this table the overwhelming predominance of the $58.6$-d contribution, i.e. the rotational period. As expected, it is excited by the proximity of the $1:1$ secondary resonance with the proper frequency $\omega_z$. We can see that the amplitude associated is roughly the mean value of $J_c$ as can be read from Figure \ref{fig:polarcore} ($858.344$ arcsec = $14.306$ arcmin). We also note some similarities with the precession of the rotation axis of the mantle (Table \ref{tab:polarmantle}), the frequencies involved being the same ones.

\section{Conclusion}

\par In this paper we have investigated the 4-degree of freedom behaviour of a rotating Mercury composed of a rigid mantle and a fluid ellipsoidal core, using both analytical and numerical tools with good agreement. We have emphasized the influence of the proximity of a resonance with the spin of Mercury, that can raise the velocity field of the fluid constituting the core. We cannot exclude a possibility of indirect detection of this effect by measuring Mercury's magnetic field. We have also shown the variations of the behaviour of the obliquity of the mantle with respect to the polar flattening of the core, this flattening being linked with the distance of the system from the resonance. These variations should be negligible unless the core is trapped into the resonance with the spin of Mercury. We have also shown that neither the observations of the longitudinal motion of Mercury, nor of its polar one, could be inverted to get the shape of the core. However, they will give information on its size (i.e. the parameter $\delta$).

\par Future works should take the viscosity of the fluid into account. It is assumed to alter the response of the core of Mercury to slow (i.e. $\approx10^5$-year period) excitations, the planet being therefore assumed as rigid. As a consequence, a study of the spectral response of the rotation of Mercury on periodic solicitations with respect to the viscosity is worth studying.

\section*{acknowledgements}

This study benefited from the financial support of the contract Prodex C90253 ``ROMEO'' from BELSPO. Beno\^it Noyelles was also supported by the Agenzia Spaziale Italiana (ASI grant "Studi di Esplorazione del Sistema Solare"), and thanks Alessandra Celletti and Luciano Iess for their reception in Rome. We also thank Nicolas Rambaux, Marie Yseboodt and Tim Van Hoolst for fruitful discussions.

\label{lastpage}

\end{document}